\documentclass[12pt]{iopart}
\usepackage{cite}
\expandafter\let\csname equation*\endcsname\relax
\expandafter\let\csname endequation*\endcsname\relax
\usepackage{amsmath,amssymb,amsfonts}
\usepackage{algorithmic}
\usepackage{graphicx}
\usepackage{textcomp}
\usepackage{bm}
\usepackage{subfigure}
\usepackage{amsfonts}

\usepackage{amssymb}
\usepackage{amsmath}
\usepackage{epstopdf}
\usepackage{color}
\usepackage{threeparttable}
\usepackage{xfrac}
\usepackage{dsfont}
\usepackage{booktabs}
\usepackage{multirow}

\newcommand{\ie}{\textit{i.e., }}
\newcommand{\myetal}{\textit{et al. }}
\newcommand{\eg}{\textit{e.g., }}

\newcommand{\vct}[1]{\boldsymbol{#1}}
\newcommand{\newMat}[1]{\mathbf{#1}}
\begin{document}

\title[GSQAS: Graph Self-supervised Quantum Architecture Search]{GSQAS: Graph Self-supervised Quantum Architecture Search}

\author{Zhimin He$^1$, Maijie Deng$^2$, Shenggen Zheng$^3$, Lvzhou Li$^4$ and Haozhen Situ$^{5*}$}
\address{$^1$School of Electronic and Information Engineering, Foshan University, Foshan 528000, China}
\address{$^2$School of Mechatronic Engineering and Automation, Foshan University, Foshan 528000, China}
\address{$^3$Peng Cheng Laboratory, Shenzhen, 518055 China}
\address{$^4$Institute of Quantum Computing and Computer Theory, School of Computer Science and Engineering, Sun Yat-Sen University, Guangzhou 510006, China}
\address{$^5$College of Mathematics and Informatics, South China Agricultural University, Guangzhou 510642, China}

\hspace{3.6pc} *Corresponding author
\ead{situhaozhen@gmail.com }
\vspace{10pt}
\begin{indented}
\item[]March 2023
\end{indented}

\begin{abstract}
Quantum Architecture Search (QAS) is a promising approach to designing quantum circuits  for variational quantum algorithms (VQAs).
However, existing QAS algorithms require to evaluate a large number of quantum circuits during the search process, which makes them computationally demanding and limits their applications to large-scale quantum circuits.
Recently, predictor-based QAS has been proposed to alleviate this problem by directly estimating the performances of circuits according to their structures with a predictor trained on a set of labeled quantum circuits.
However, the predictor is trained by purely supervised learning, which suffers from poor generalization ability when labeled training circuits are scarce.
It is very time-consuming to obtain a large number of labeled quantum circuits because the gate parameters of quantum circuits need to be optimized until convergence to obtain their ground-truth performances.
To overcome these limitations, we propose GSQAS, a graph self-supervised QAS, which trains a predictor based on self-supervised learning.
Specifically, we first pre-train a graph encoder on a large number of unlabeled quantum circuits using  a well-designed pretext task in order to generate meaningful representations of circuits.
Then the downstream predictor is trained on a small number of quantum circuits' representations and their labels.
Once the encoder is trained, it can apply to different downstream tasks.
In order to better encode the spatial topology information and avoid the huge dimension of feature vectors for large-scale quantum circuits, we design a scheme to encode quantum circuits as graphs.
Simulation results on searching circuit structures for variational quantum eigensolver and quantum state classification show that GSQAS outperforms the state-of-the-art predictor-based QAS, achieving better performance with fewer labeled circuits.
\end{abstract}

%
\vspace{2pc}
\section{Introduction}
Variational quantum algorithms (VQAs) have attracted much attention in the noisy intermediate-scale quantum (NISQ) era due to their resilience to errors and high flexibility to the limitations of NISQ devices such as limited qubits connectivity and shallow depth
\cite{peruzzo2014variational,cerezo2021variational,higgott2019variational,kandala2017hardware,jones2019variational,moussa2020quantum}.
VQAs have been successfully applied to a plethora of applications, \eg finding the ground and excited states \cite{higgott2019variational,jones2019variational,kandala2017hardware}, dynamical quantum simulation \cite{mcardle2019variational,yao2021adaptive}, quantum compiling \cite{he2021variational,khatri2019quantum}, combinatorial optimization \cite{farhi2014quantum,moussa2020quantum} and Hamiltonian learning \cite{shi2022parameterized}. The applications of VQAs have great potential in the demonstration of quantum advantage in NISQ era \cite{beer2020training,li2020quantumCNN,niu2022entangling,situ2020quantum,shi2023quantum}.

VQAs are implemented by parameterized quantum circuits (PQCs) running on the NISQ devices. The parameters of PQC are optimized to minimize the objective function by a classical optimizer.
A bottleneck of VQAs is the trainability problem, \ie the barren plateau phenomenon \cite{mcclean2018barren,marrero2021entanglement}, which refers to the phenomenon that the gradients vanish exponentially with the problem size.
Current researches show that well-designed PQCs can avoid the barren plateau problem and speed up the training of VQAs \cite{cerezo2021cost,mcclean2018barren,pesah2021absence,sharma2022trainability}.

To enhance the trainability and expressibility of PQCs, a systematic and automatic strategy has been proposed to design a problem-specific PQC for a given VQA \cite{meng2021quantum,chivilikhin2020mog,cincio2021machine,grimsley2019adaptive,he2021variational,li2020quantum,zhang2022differentiable,zhang2021neural}.
The automatic design of PQCs is known as quantum architecture search \cite{du2022quantum,he2022AQT,he2022search,zhang2021neural,zhang2022differentiable,linghu2022quantum,lu2021markovian}, ansatz architecture search \cite{li2020quantum}, structure optimization for PQCs \cite{ostaszewski2021structure}, adaptive variational algorithm \cite{grimsley2019adaptive} and circuit learning \cite{cincio2021machine} in different literatures.
In this paper, we follow the term \emph{quantum architecture search} (QAS) as it is the most widely used one.
QAS algorithms are composed of a search module and an evaluation module. The search module uses a search strategy to explore high-performance quantum circuits in the search space while the evaluation module calculates the performance of quantum circuits as feedback to guide the search module.

Most of the current QAS algorithms focused on proposing different search strategies to improve performance, \eg evolutionary algorithms \cite{chivilikhin2020mog,huang2022robust,rattew2019domain}, gradient-based algorithms \cite{zhang2022differentiable} and reinforcement learning \cite{zhang20,he2021variational,ostaszewski2021reinforcement,kuo2021quantum}.
Although QAS can provide high-performance PQCs for VQAs and largely reduce the number of demanded quantum gates and the circuit depth, the computational complexity is the main constraint as QAS algorithms require to search and evaluate a large number of quantum circuits.
The calculation of ground-truth performance is very time-consuming as the gate parameters are optimized until convergence.

Predictor-based QAS (PQAS) is an effective strategy to reduce the computational complexity of QAS.
Zhang \myetal \cite{zhang2021neural} trained a predictor on a set of quantum circuits with ground-truth performance labels and directly estimated the performances of quantum circuits by the trained predictor without any optimization of gate parameters.
However, the predictor is trained by purely supervised learning, which suffers from the over-fitting problem when the labeled training data is scarce, leading to poor generalization.
As the acquirement of a quantum circuit's label is time-consuming, the key of PQAS is how to train a predictor with good generalization ability using a small number of labeled circuits.

In the field of machine learning, self-supervised learning (SSL) is proposed to alleviate the heavy reliance of supervised learning on labeled data.
SSL first extracts informative knowledge and learns a meaningful representation for samples from a large amount of unlabeled data by a well-designed pretext task.
Then the learned representation is used in the downstream supervised task.
As described by Turing Award winners Yoshua Bengio and Yann LeCun, SSL is the key to human-level intelligence and has achieved great success in different fields.
In this paper, we propose GSQAS, a graph self-supervised QAS, which can train an accurate predictor with much fewer labeled quantum circuits.
Our main contributions are summarized as follows.
\begin{itemize}
\item We propose to learn informative representations of quantum circuits on a large number of unlabeled circuits via graph self-supervised learning, which facilitates the training of the predictor and significantly reduces the number of required labeled circuits, leading to much less computational cost on the calculation of circuits' ground-truth performances.
\item We denote the quantum circuit structure with graph encoding, which is particularly suitable for graph self-supervised learning. The graph encoding has no restriction on the two-qubit gates and can represent any quantum circuit.  The node vector is composed of a type vector and a position vector, which can better encode the gate type information and spatial information, and avoid huge dimensions of the feature vectors for large-scale quantum circuits.
\item 
    Simulations on variational quantum eigensolver and quantum state classification show that the proposed methods achieve better performance using fewer labeled circuits than the state-of-the-art predictor-based QAS.
\end{itemize}

The rest of the paper is organized as follows. Section 2 describes the terminologies and notations. Section 3 introduces the related work, including quantum architecture search and self-supervised learning. The proposed graph self-supervised quantum architecture search is described in Section 4. Simulation results on variational quantum eigensolver and variational quantum classifier are shown in Section 5. We draw a conclusion and discuss possible future work in Section 6.

\section{Terminologies and notations}
We illustrate the related terminologies and notations in Table \ref{tab:terminolygy} for better understanding.
\begin{table}
\centering
\caption{The related terminologies and notations.}
\label{tab:terminolygy}
\setlength{\tabcolsep}{3pt}
\centering
\begin{tabular}{lp{4in}}
\hline
Term& Description \\
\hline
Circuit training & Optimizing the gate parameters in the circuit until convergence.\\
Ground-truth performance & The performance after circuit training.\\
Label of a quantum circuit& The ground-truth performance of the quantum circuit.\\
Predicted performance & The performance estimated by the predictor.\\
Training samples/circuits& Labeled circuits used to train a predictor.\\
Screening circuits& Circuits required to obtain their predicted performances in the large-scale screening step in Figure \ref{Fig:predictorQAS}.\\
Candidate circuits& Circuits required to obtain their ground-truth performances in the final selection step in Figure \ref{Fig:predictorQAS}.\\
$n$ & Number of qubits.\\
$N$ & Number of nodes/gates of a graph/circuit.\\
$F$ & Dimension of the feature vector of a node.\\
$l$ & Dimension of the latent representation of a node.\\
$\vct x_i \in \mathbb{R}^{F}$ & Feature vector of node $i$.\\
$\newMat{X} = [\vct x_1, ..., \vct x_N]^T$ & Feature matrix of a graph.\\
$\newMat{A}\in \mathbb{R}^{N \times N}$ & Adjacency matrix of a graph.\\
$\vct z_i \in \mathbb{R}^{l}$ & Latent representation of node $i$\\
$\newMat{Z} = [\vct z_1, ..., \vct z_N]^T$ & Latent representation of a graph/circuit.\\
$q_{\vct \phi}$ & Encoder where $\vct \phi$ are its trainable parameters.\\
$p_{\vct \theta}$ & Decoder where $\vct \theta$  are its trainable parameters.\\
$f_{\vct \psi}$ & Predictor where $\vct \psi$  are its trainable parameters.\\
\hline
\end{tabular}
\end{table}

\section{Related works}
Our work is mainly related to quantum architecture search (QAS) and self-supervised learning (SSL).
\subsection{Quantum architecture search}
QAS aims to automatically design high-performance quantum circuits for given tasks, \eg state preparation, unitary compilation, error correction, state classification, variational quantum eigensolver, and quantum approximate optimization algorithm.
QAS requires to search and evaluate a large number of quantum circuits.
However, the evaluation of a quantum circuit is very time-consuming as it requires multiple derivations of the parameters with respect to each quantum gate.
The computational complexity of current QAS algorithms is unaffordable for large-scale quantum circuits.

Some search strategies have been proposed to accelerate QAS algorithms.
Zhang \myetal relaxed the discrete search space of quantum circuits onto a continuous domain by a probabilistic model and optimized the structure by gradient descent \cite{zhang2022differentiable}.
A metaQAS algorithm is proposed to use prior experiences instead of searching the circuit structure from scratch for each task \cite{he2022AQT}.
Initialization heuristics of the structure and gate parameters are learned from a set of training tasks and embed the across-task knowledge.
With the initialization heuristics, the QAS algorithm converges after a small number of updates in new tasks.
Another strategy to accelerate QAS algorithms is to use the approximate performances of quantum circuits as feedback instead of the ground-truth performances.
A gate-sharing technique is used to estimate the approximate performances \cite{wang2022quantumnas}.
The performances of quantum circuits are estimated directly with the gate parameters inherited from the trained super circuit which integrates all the candidate circuit structures, instead of optimizing the gate parameters of each quantum circuit from scratch.
The key assumption of the gate-sharing technique is that when the circuit is evaluated by inherited parameters, the ranking of circuits is relatively consistent with the ranking obtained from training them independently.
However, the validity of this assumption depends on the search space, the training of the super circuit and VQA tasks, which cannot be guaranteed.

Predictor-based QAS (PQAS) is another effective strategy to accelerate QAS algorithms. A predictor is first trained on a small number of quantum circuits as well as their ground-truth performances (\ie labels) and then provides a curated list of promising quantum circuits for final selection.
The workflow of predictor-based QAS is shown in Figure 1.
Zhang \myetal \cite{zhang2021neural} proposed a predictor based on recurrent neural network (RNN) to predict the performances of quantum circuits and showed that predictor-based QAS could largely decrease the computational complexity and discover high-performance quantum circuits for different VQAs, \ie variational quantum eigensolver and  image  classification.
However, their image representation of quantum circuits has two restrictions, \ie two-qubit gates must be symmetric and are only permitted to act on adjacent qubits, which largely limits its applications.
\begin{figure*}
\centering
\includegraphics[width=0.8\textwidth]{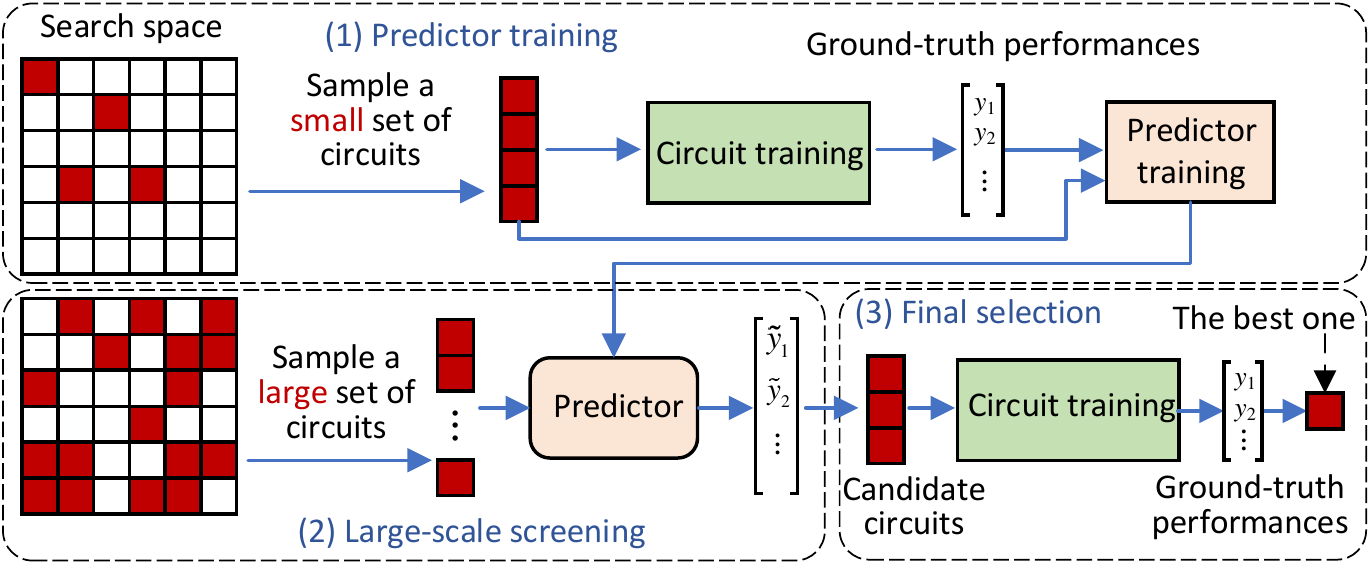}
\caption{The workflow of predictor-based QAS. (1)Predictor training: train a predictor using a small set of circuits as well as their ground-truth performances.
The ground-truth performances are obtained by circuit training.
(2)Large-scale screening: sample a large number of quantum circuits from the search space and estimate their performances by the trained predictor.
(3)Final selection: select the top-K best circuits according to the predicted performances as candidate circuits, calculate their ground-truth performances, and output the top-1 best circuit according to the ground-truth performances.
}
\label{Fig:predictorQAS}
\end{figure*}

As it is time-consuming to obtain the ground-truth performance of a quantum circuit, the crucial problem of PQAS is how to train an accurate predictor with limited labeled circuits.
A potential strategy is to extract informative knowledge from a large number of unlabeled data to facilitate the training of the predictor.

\subsection{Self-supervised learning}
Traditional deep neural networks (DNNs) heavily rely on the amount of labeled training data and suffer from poor generalization when labeled training data is scarce.
When labeled samples are limited, self-supervised learning (SSL) is emerging as a new paradigm to extract informative knowledge from a large amount of unlabeled data through well-designed auxiliary tasks (so-called pretext tasks).  The pretext task automatically provides self-defined pseudo labels as supervision for the learning of feature representations without any manual annotation.
SSL has become a promising and trending learning paradigm as it can avoid the high cost of annotating large-scale datasets and is essential in fields where the annotation is costly.
SSL has achieved great success in computer vision (CV) \cite{he2020momentum,chen2020simple,wei2022masked,he2022masked} and natural language processing (NLP) \cite{devlin2018bert,yang2019xlnet,qian2022contentvec,choi2021neural}.

Following the great success of SSL on CV and NLP, more and more SSL algorithms have been devised for graph-structured data, which is ubiquitous in various areas, \eg e-commerce, chemistry and traffic.
Current researches have proven the effectiveness of SSL for learning informative representations from unlabeled graph data, which achieves good performance \cite{velickovic2019deep,hassani2020contrastive} and generalization capability \cite{hu2020gpt,qiu2020gcc,hu2019strategies} on downstream tasks.
SSL can learn generalized feature representations from unlabeled graph data by a pretext task, which can achieve higher performance on various downstream tasks, \eg node classification\cite{wang2021self,jiang2021pre,lee2022augmentation} and graph classification \cite{luo2022clear,hassani2020contrastive,suresh2021adversarial,lin2022prototypical}.

\section{Graph self-supervised quantum architecture search (GSQAS)}
\begin{figure*}
\centering
\includegraphics[width=0.9\textwidth]{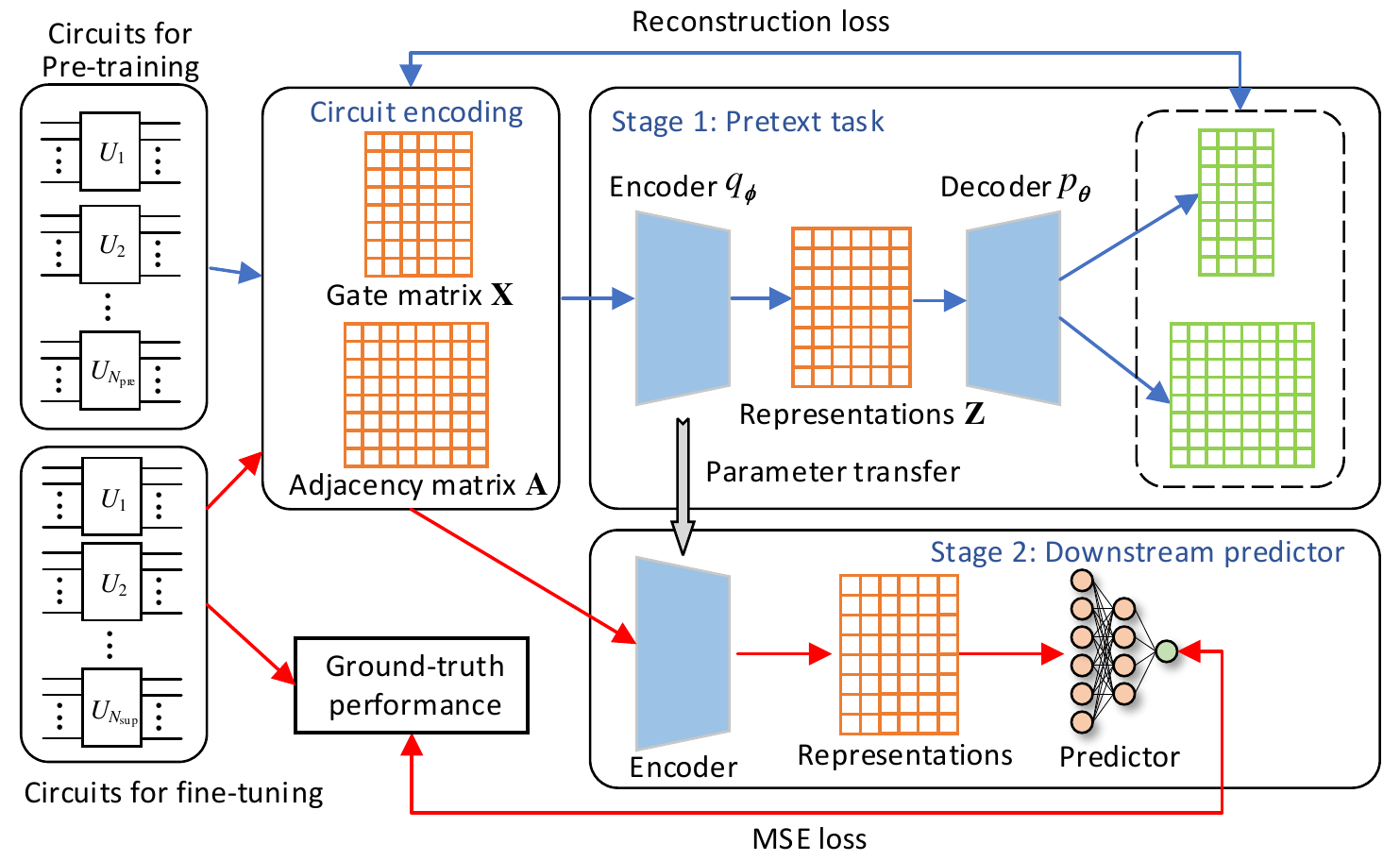}
\caption{The predictor training part of GSQAS. Stage 1: unsupervised learning of a pretext task using a large number of unlabeled circuits. Stage 2: fine-tuning the encoder and the predictor using a small number of labeled circuits. The blue lines indicate the workflow of Stage 1 while the red lines denote the workflow of Stage 2.
}
\label{Fig:ssl}
\end{figure*}

In this section, we describe the proposed GSQAS, which incorporates SSL and graph encoding for quantum circuits to facilitate the training of the predictor.
Figure \ref{Fig:ssl} provides the training process of the predictor in GSQAS.
In order to find a good initialization for the encoder in the downstream prediction task, the encoder is trained on a large number of unlabeled quantum circuits in a pretext task.
In the pretext task, an encoder $q_{\vct \phi}$ first embeds the quantum circuit into a latent representation $\newMat{Z}$, and then a decoder $p_{\vct \theta}$ tries to recover the original quantum circuit from $\newMat{Z}$. As a better reconstruction indicates richer information integrated into $\newMat{Z}$,
the loss function $\mathcal{L}_{pre}$ of the pretext task is defined as the reconstruction loss, aiming to minimize the difference between the reconstructed and the original quantum circuits.
\begin{equation}
\vct \phi_{pre},\vct \theta_{pre}  = \mathop{\arg \min} \limits_{\vct \phi, \vct \theta} \mathcal{L}_{pre}(q_{\vct \phi},p_{\vct \theta},\mathcal{D}_{pre}),	
\label{eq:encoderEB}
\end{equation}
where $\vct \phi$ and $\vct \theta$ are trainable parameters of the encoder and the decoder, respectively. $\mathcal{D}_{pre}$ is the training set of quantum circuits in the pretext task.

After the pre-training stage, the pre-trained encoder $q_{\vct \phi_{pre}}$ is combined with a predictor $f_{\vct \psi}$, where $\vct \psi$ are the trainable parameters of the predictor.
The encoder and the predictor are fine-tuned using a small number of  quantum circuits $\mathcal{D}_{sup}$ as well as their ground-truth performances $\vct y_{sup}$
\begin{equation}
\vct \phi^*, \vct \psi^* = \mathop{\arg \min} \limits_{\vct \phi, \vct \psi} \mathcal{L}_{sup}(q_{\vct \phi},f_{\vct \psi},\mathcal{D}_{sup},\vct y_{sup}),
\label{eq:encoderEB}
\end{equation}
where the parameters $\vct \phi$ of the encoder are initialized with $\vct \phi_{pre}$.

The proposed circuit encoding scheme is first introduced in Section \ref{sec:encoding}. We describe the pretext task learning and downstream predictor learning in Section \ref{sec:pretextTask} and Section \ref{sec:DSPredictor}, respectively.

\subsection{Circuit encoding scheme}
\label{sec:encoding}
\begin{figure}
\centering
\includegraphics[width=0.7\textwidth]{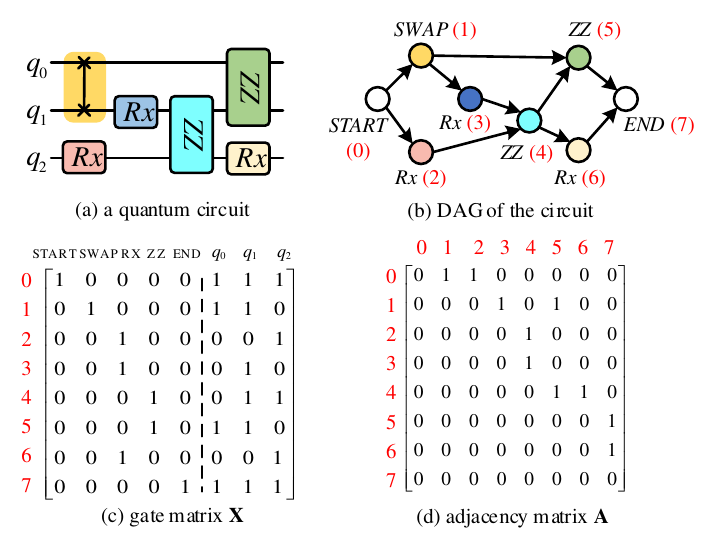}
\caption{An example of the graph encoding scheme.}
\label{Fig:circuitEncoding}
\end{figure}

Instead of the image encoding in Ref. \cite{zhang2021neural}, we propose a graph encoding scheme to represent quantum circuits.
A quantum circuit with $N$ gates can be represented by a directed acyclic graph $\mathcal{G}= (\mathcal{V},\mathcal{E})$ with $N$ nodes, where $\mathcal{V}$ and $\mathcal{E}$ denote a set of nodes and edges, respectively. Each node $v_i \in \mathcal{V}$ is associated with a gate operation and denoted by an $F$-dimensional feature vector $\vct x$.
We add a \textit{START} node and an \textit{END} node to mimic the input and output of the circuit.
In order to avoid the huge dimension of the feature vector for large-scale quantum circuits, we use a one-hot vector $\vct x^{(t)}$ to denote the type of quantum gate and an $n$-dimensional position vector $\vct x^{(q)}$ to describe the qubit(s) the gate acts on, where $n$ is the number of qubits.
Another advantage of encoding the quantum gate with a type vector and a position vector is that the quantum gates with
the same type or operated on the same qubit have similar encoding. It can better preserve the gate type information and position information.
We concatenate $\vct x^{(t)}$ and $\vct x^{(q)}$ to form the feature vector $\vct x$ of the node, \ie $\vct x = [\vct x^{(t)}, \vct x^{(q)}]$.
The set of nodes is denoted by a gate matrix $\newMat{X} = [\vct x_1, ..., \vct x_N]^T$.
The set of edges is denoted by an adjacency matrix $\newMat{A}\in \mathbb{R}^{N \times N}$, describing the connection of quantum gates, where $\newMat{A}_{ij}= \mathds{1}[(v_i, v_j) \in \mathcal{E}]$ for $0\leq i,j<N$.
Compared to image encoding \cite{zhang2021neural}, the proposed graph encoding has no restriction on the two-qubit gates and can represent any quantum circuit.
The gate matrix and adjacency matrix well preserve the gate information and the topology information of the quantum circuit. Figure \ref{Fig:circuitEncoding} illustrates an example of the graph encoding.

\subsection{Pretext task learning}
\label{sec:pretextTask}

In the pretext task, we aim to learn an encoder to provide informative representations for the downstream predictor. Because the training of the pretext task uses unlabeled quantum circuits, it is not related to a specific task. Once the encoder is learned, it can apply to different downstream tasks, \eg VQE, QAOA and quantum classifiers.

Graph reconstruction provides a natural self-supervision pretext task for the training of graph-based predictors.
It exploits both the structure information and the node features associated with the nodes in a graph, which significantly improves the predictive performance.
As shown in Figure \ref{Fig:ssl}, the self-supervision model consists of an encoder and a decoder.
The encoder embeds the graph data encoded in the gate matrix $\newMat{X}$ and the adjacency matrix $\newMat{A}$ into a continuous representation, while the decoder reconstructs $\newMat{X}$ and $\newMat{A}$ from the representation.

We use variational graph autoencoders (VGAE) \cite{kipf2016variational}, which has shown promising performance in learning latent representations for graph, to implement the encoder and the decoder. In order to learn a latent representation that is invariant to isomorphic graphs, graph isomorphism networks (GIN) \cite{KeyuluXu2018HowPA} are used to map the quantum circuit encoding to its latent representation instead of graph convolutional networks in VGAE \cite{kipf2016variational}.

\subsubsection{Encoder}
The encoder embeds the node feature information and structure information  to derive latent representations $\newMat{Z} = [\vct z_1, ..., \vct z_N]^T$ of a graph, where $\vct z_i $ is a $l$-dimensional representation of node $i$.
The encoder regards the representations of nodes as Gaussian random variables.
As finding the posterior distribution $q_{\vct \phi}(\newMat{Z}|\newMat{X},{\newMat{A}})$ is difficult and intractable, it is approximated by a variational distribution \cite{kipf2016variational}
\begin{equation}
	q_{\vct \phi}(\newMat{Z}|\newMat{X},{\newMat{A}}) = \prod_{i=1}^Nq_{\vct \phi}(\vct z_i | \newMat{X},{\newMat{A}}).
	\label{eq:encoder}
\end{equation}
$\vct z_i$ is approximated by a Gaussian distribution, \ie  $q_{\vct \phi}(\vct z_i | \newMat{X},{\newMat{A}}) = \mathcal{N}(\vct z_i|\vct \mu_i,\text{diag}(\vct \sigma^2_i))$.
$\vct \mu_i$ and $\vct \sigma_i$ are $l$-dimensional  mean and standard deviation vectors corresponding to node $i$, which are modeled by graph neural networks.
 Graph isomorphism networks (GIN) are used to model the mean and standard deviation
\begin{equation}
	\vct \mu = \text{GIN}_{\vct \mu}(\newMat{X}, \widehat{\newMat{A}}), \log \vct \sigma = \text{GIN}_{\vct \sigma}(\newMat{X},\widehat { \newMat{A}}).
\end{equation}
where $\widehat{\newMat{A}} = \newMat{A} + \newMat{A}^T$ transfers original directed graphs into undirected graphs in order to use bi-directional information.
We use an $L$-layer GIN to generate the node embedding matrix $\newMat{H}$. The embedding matrix in the $k$-th layer is defined as
\begin{equation}
	\newMat{H}^{(k)} = \text{MLP}^{(k)}((1+\varepsilon^{(k)})\cdot \newMat H^{(k-1)} + \widehat{\newMat{A}}\newMat H^{(k-1)}), k = 1,2,...,L,
	\label{eq:encoderEB}
\end{equation}
where $\newMat H^{(0)} = \newMat X$ and MLP denotes a multi-layer perception with the Linear-Batchnorm-ReLU structure. $\varepsilon$ is a learnable bias.
$\newMat H^{(L)}$ is fed into two fully-connected layers to get $\vct \mu =[\vct \mu_1,...,\vct \mu_N]^T$ and $\vct \sigma=[\vct \sigma_1,...,\vct \sigma_N]^T$, respectively.
With the reparameterization trick \cite{DiederikPKingma14}, we can sample from the variational posterior $\vct z_i \sim q_{\vct \phi}(\vct z_i | \newMat{X},{\newMat{A}})$ as
\begin{equation}
    \vct z_i = \vct \mu_i + \vct \sigma_i \odot \vct \epsilon; \vct \epsilon \sim \mathcal{N}(0,\newMat I).
\end{equation}
where $\odot$ denotes element-wise multiplication.

\subsubsection{Decoder}
The decoder is a generative model aiming to reconstruct the adjacency matrix ${\newMat{A}}$ and the feature matrix $\newMat X =\text{concat}(\newMat X^{(t)}, \newMat X^{(q)})$ from the latent representation $\newMat{Z}$.
The decoder can be formulated as
\begin{equation}
\begin{aligned}
	&p_{\vct \theta}({\newMat A}|\newMat Z) = \prod_{i=1}^N\prod_{i=j}^Np_{\vct \theta}({\newMat A}_{ij} | \newMat Z), \\
&   \text{with } p_{\vct \theta}({\newMat A}_{ij} =1|\newMat Z)=\text{sigmoid}(\newMat W_a (\newMat Z\newMat Z^T)_i + \vct b_a)_{j},
	\label{eq:encoderA}
\end{aligned}
\end{equation}
\begin{equation}
\begin{aligned}
	p_{\vct \theta}({\newMat X}^{(t)}|\newMat Z)& = \prod_{i=1}^N p_{\vct \theta}(\vct x_i^{(t)}| \vct z_i) \\
& = \prod_{i=1}^N \text{softmax}(\newMat W_t \vct z_i + \vct b_t)_{k_i},
	\label{eq:encoderXt}
\end{aligned}
\end{equation}
\begin{equation}
\begin{aligned}
&	p_{\vct \theta}({\newMat X}^{(q)}|\newMat Z) = \prod_{i=1}^N \prod_{j=1}^n p_{\vct \theta}({\newMat X}^{(q)}_{ij}| \vct z_i), \\
& \text{with }  p_{\vct \theta}({\newMat X}^{(q)}_{ij} =1|\vct z_i)=\text{sigmoid}(\newMat W_q \vct z_i + \vct b_q)_{j},
\label{eq:encoderEBXq}
\end{aligned}
\end{equation}
where $\newMat W_a, \newMat W_t, \newMat W_q$ are  the trainable weights and  $\vct b_a, \vct b_t, \vct b_q$ are the trainable biases of the decoder.
$\vct x^{(t)}_i$ is a one-hot vector, which indicates the type of the selected quantum gate at the $i$-th node and $k_i$ denotes the index of the selected quantum gate,
\ie $k_i = \mathop{\arg \max} \limits_{j} {\newMat X}^{(t)}_{ij}  $.

\subsubsection{Pretext training}
The model weights of the encoder and the decoder are trained by iteratively maximizing a tractable variational lower bound
\begin{equation}
\begin{aligned}
\mathcal{L} =  \mathbb{E}_{q_{\vct \phi}(\newMat{Z}|\newMat{X},{\newMat{A}})}[\log p_{\vct \theta}({\newMat X}^{(t)}, {\newMat X}^{(q)}, {\newMat A}|\newMat Z) ] - KL(q_{\vct \phi}(\newMat Z|\newMat X, {\newMat A})\|p(\newMat Z)),
\label{eq:encoderEBXq}
\end{aligned}
\end{equation}
where $p_{\vct \theta}({\newMat X}^{(t)}, {\newMat X}^{(q)}, {\newMat A}|\newMat Z) = p_{\vct \theta}({\newMat X}^{(t)}|\newMat Z)p_{\vct \theta}({\newMat X}^{(q)}|\newMat Z)p_{\vct \theta}({\newMat A}|\newMat Z)$ as we assume that ${\newMat X}^{(t)}, $${\newMat X}^{(q)}, $$ {\newMat A}$  are conditionally independent given $\newMat Z$.
The first term is an expected negative reconstruction error while the second term acts as a regularizer where $KL$ denotes the Kullback-Leibler divergence of the approximate posterior $q_{\vct \phi}(\newMat Z|\newMat X, {\newMat A})$ from the prior $p(\newMat Z)$.
We use a Gaussian prior, \ie $p(\newMat Z) = \prod_i\mathcal{N}(\vct z_i|0, \newMat I)$.
The loss is optimized by mini-batch gradient descent over the model weights of the encoder and the decoder.

\subsection{Downstream predictor learning}
\label{sec:DSPredictor}
\begin{figure}
\centering
\includegraphics[width=0.5\textwidth]{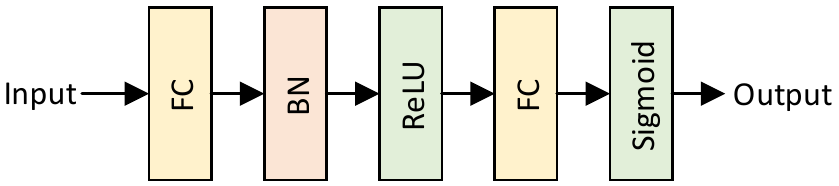}\\
\caption{The architecture of the predictor.}
\label{Fig:rg}
\end{figure}

In the downstream task, the encoder takes a graph as input and outputs the latent representations of all nodes.
The predictor takes the average of all nodes' representations as input.
The architecture of the encoder is the same as the one in the pretext task.
The parameters of the encoder inherit from the pre-trained encoder in the pretext task.
The predictor is a multi-layer perception with two fully connected layers. 
The activation function of the first and the second layers are ReLU and Sigmoid, respectively.
We add a batch normalization layer between the first linear layer and the ReLU layer in order to accelerate training and improve generalization
capability. Figure \ref{Fig:rg} shows the architecture of the predictor. In this paper, we consider two self-supervised training schemes, \ie unsupervised representation learning denoted by GSQAS(URL) and pre-training and fine-tuning denoted by GSQAS(PF).
In GSQAS(URL), the parameters of the encoder are frozen and only the parameters of the predictor are optimized in the downstream task.
In GSQAS(PF), the pre-trained parameters in the pretext task are leveraged as the initial parameters for the encoder. The parameters of both the encoder and the predictor are optimized.
Mean square error (MSE) is adopted as the loss function and the parameters are optimized by mini-batch gradient descent.

\section{Results}
We verify the performance of the proposed GSQAS on two specific VQA tasks, \ie variational quantum eigensolver (VQE)  for transverse field Ising model (TFIM) and variational quantum classifier (VQC) for quantum state classification.
The VQE energy and the classification accuracy are used as the label of a quantum circuit in the VQE and VQC tasks, respectively.

We first compare the proposed graph encoding scheme (GQAS) with the predictor-based QAS based on image
encoding (PQAS) \cite{zhang2021neural}  to verify the performance of the graph encoding scheme.
In GQAS, we use the average of all nodes' feature vectors as the input of the predictor. The architecture of the predictor in GQAS is the same as the one in GSQAS as shown in Figure \ref{Fig:rg} in order to make a fair comparison.
Mean square error (MSE) is adopted as the loss function and the predictor is optimized by mini-batch gradient descent.

The GSQAS using an unsupervised representation learning scheme is denoted by GSQAS(URL) while the one using a pre-training and fine-tuning scheme is denoted by GSQAS(PF).
Compared with GQAS, GSQAS(URL) and GSQAS(PF) have an additional encoder that is trained on unlabeled quantum circuits in the pretext task. Thus, GQAS is the supervised counterpart of GSQAS(URL) and GSQAS(PF).
As the parameters of the encoder are frozen in the downstream task, the number of trainable parameters in GSQAS(URL) is the same as its supervised counterpart GQAS.
All numerical simulations are implemented on TensorCircuit \cite{zhang2022tensorcircuit}.

\subsection{Variational quantum eigensolver for TFIM}
In this section, the proposed algorithm is used to accomplish the VQE tasks by finding quantum circuits to achieve the ground-state energy of a 6-qubit TFIM. The Hamiltonian of the system is denoted by
\begin{equation}
	H = \sum_{i=0}^5\sigma_z^i\sigma_z^{{(i+1)} \text{ mod } 6} + \sigma_x^i.
	\label{eq:HamiltonianTFIM}
\end{equation}
The ground-state energy $E_0$ of the TFIM with periodic boundary condition is -7.7274066.
We follow the search space of Ref. \cite{zhang2021neural}.
The primitive set of quantum gates consists of  $H$, $R_x$, $R_y$, $R_z$, $XX$, $YY$ and $ZZ$.
$H$ denotes the Hadamard gate. $R_x$, $R_y$ and  $R_z$ are rotation gates, where $R_x = e^{-i\theta \sigma_x/2}$ and similarly for $R_y$ and $R_z$. $XX$, $YY$ and $ZZ$ are parameterized two-qubit gates, where $XX_{12} = e^{-i\theta \sigma_x^1\sigma_x^2/2}$ and similarly for $YY$ and $ZZ$. The number of quantum gates and the maximum depth of the circuit are set to $36$ and $10$, respectively. According to this setting, we have $\newMat{X} \in \mathbb{R}^{38\times 15}$, $\newMat{A} \in \mathbb{R}^{38\times 38}$.
We set $l=F$ \footnote[1]{With $l=F$, the architecture of the predictor in GSQAS is the same as the one in GQAS for fair comparison.} and thus have $\newMat{Z} \in \mathbb{R}^{38\times 15}$.
Similar to Ref. \cite{zhang2021neural}, we use a layerwise pipeline to generate quantum circuits, \ie  the circuit is constructed by iteratively adding a layer of $n/2$ gates, where $n$ is the number of qubits.
More specifically, we first select a type of quantum gate and place the gate on either all the odd qubits or all the even qubits.
If the selected gate is a two-qubit gate, it acts on either qubits $(1,2),(3,4),(5,0)$ or qubits $(0,1),(2,3),(4,5)$.
As previous works show that VQE started from  $|+\rangle$  state may find a solution closer to the  ground-state energy, the quantum circuits start with a layer of $n$ Hadamard gate on all qubits. For more detail on the search space and the layerwise generation, please refer to Ref. \cite{zhang2021neural}.

\begin{figure}
\centering
\includegraphics[width=0.6\textwidth]{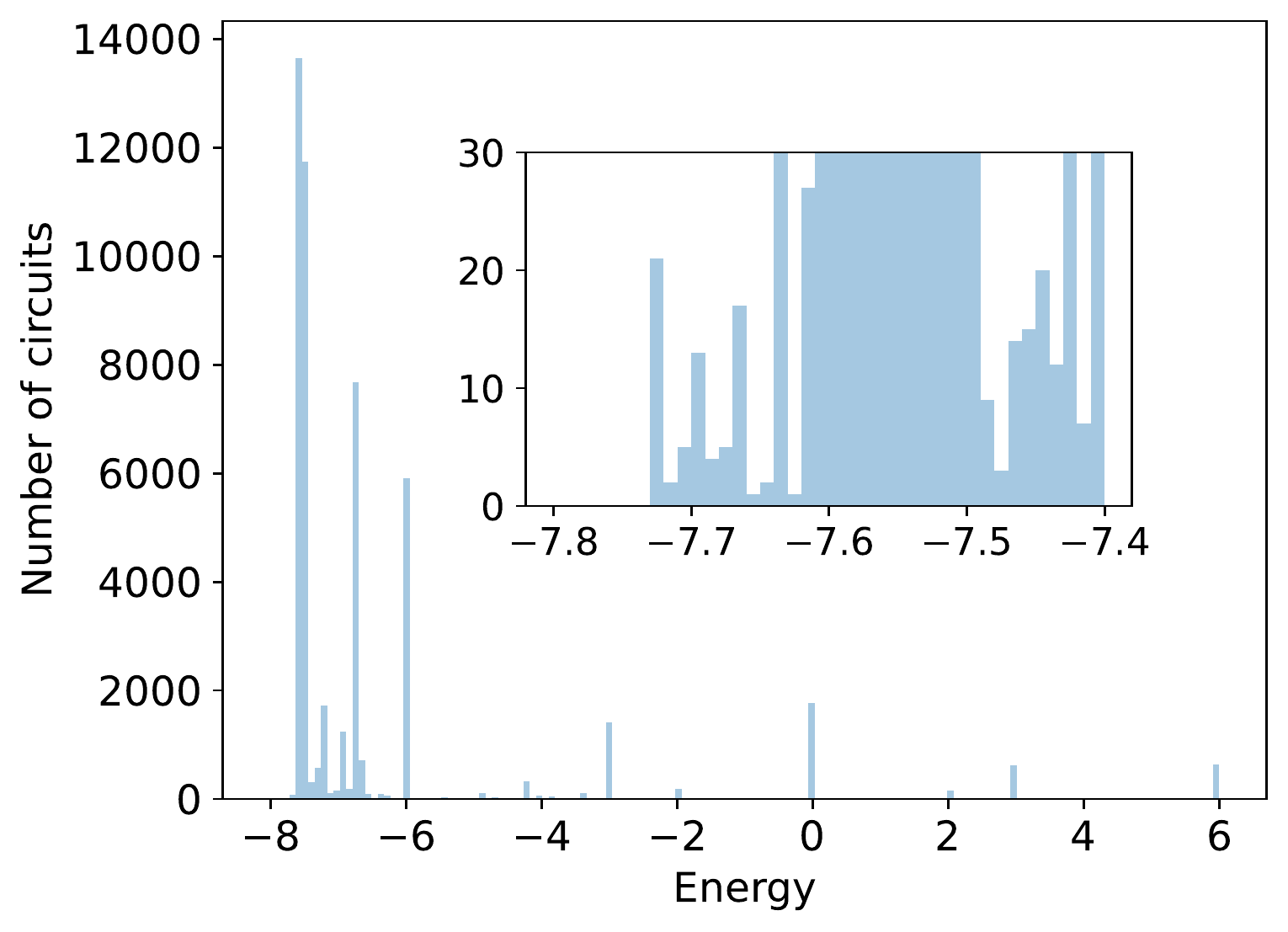}\\
\caption{Energy distribution of the TFIM model with 50,000 randomly generated quantum circuits.
The energy of a quantum circuit is calculated by optimizing the gate parameters of this circuit until convergence.
The inset shows the distribution between -7.8 and -7.4.}
\label{Fig:distribution}
\end{figure}

We randomly sample 50,000 quantum circuits according to the layerwise generation pipelines and show the VQE energy distribution of these quantum circuits in Figure \ref{Fig:distribution}. We can observe that the energy distribution is unbalanced and not smooth in the search space. It is difficult to train a single predictor to accurately predict the energies of the quantum circuits with such distribution. Similar to Ref. \cite{zhang2021neural}, we use a two-stage screening: (1) a binary classifier is first trained to filter out the quantum circuits with poor performances; (2) a predictor is trained to estimate the energy of the remained quantum circuits.
For the binary classifier, the training circuit is labeled as a good circuit ($y=1$) if it achieves energy lower than -7.55, otherwise is labeled as a poor circuit ($y=-1$). For the predictor, We use the normalized deviation of the estimated energy as the predicted label, \ie $y_i=(E-E_0)/14\in [0,1)$.
Both the binary classifier and the predictor take the output of the encoder as input. The structure of the binary classifier is the same as the predictor except binary cross-entropy loss is used instead of MSE. The binary classifier and the predictor use the same training circuits.

\subsubsection{2D visualization of the proposed circuit encoding}
\begin{figure}
\centering
\subfigure[Graph encoding]{
\includegraphics[width=0.45\textwidth]{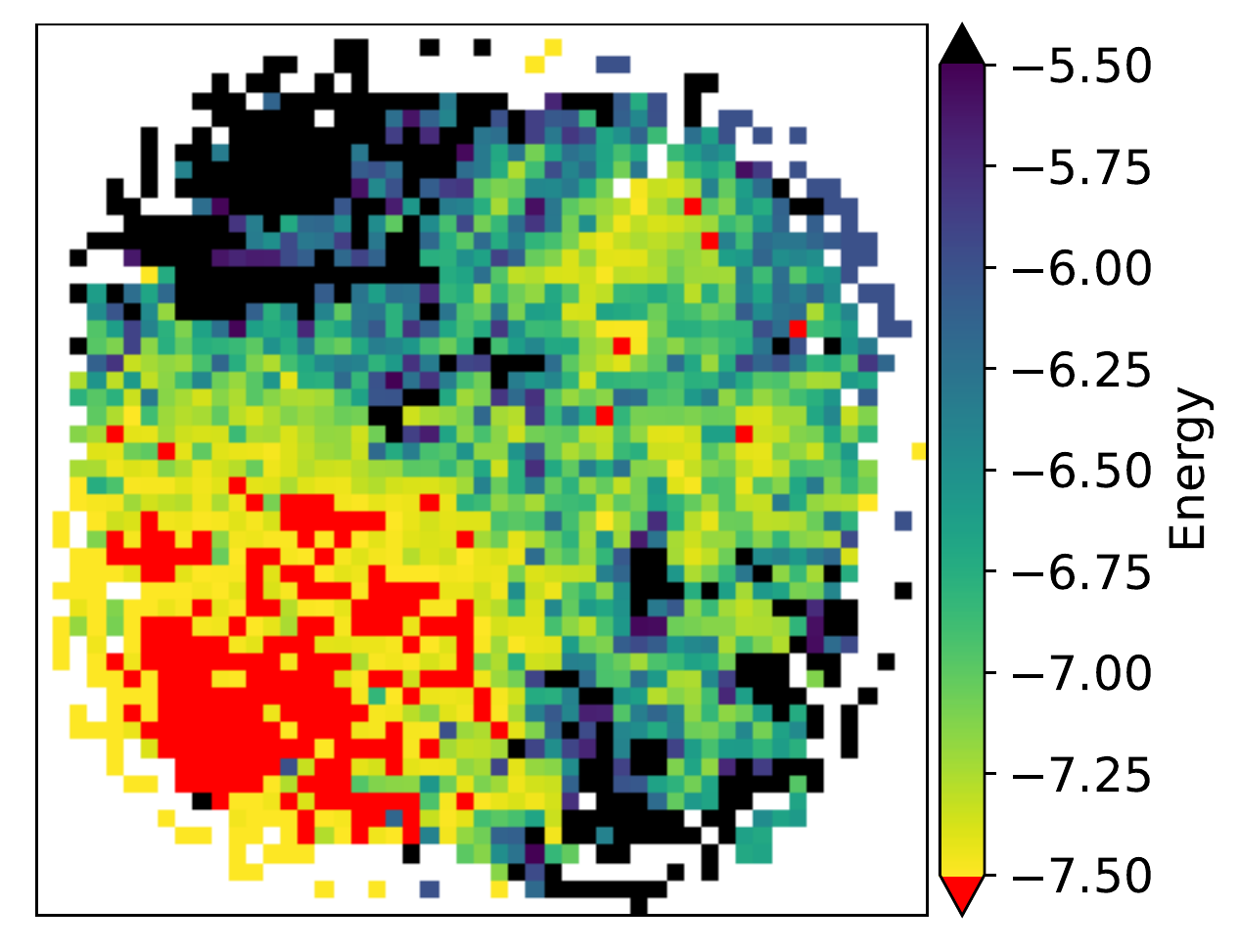}}
\subfigure[Image encoding]{
\includegraphics[width=0.45\textwidth]{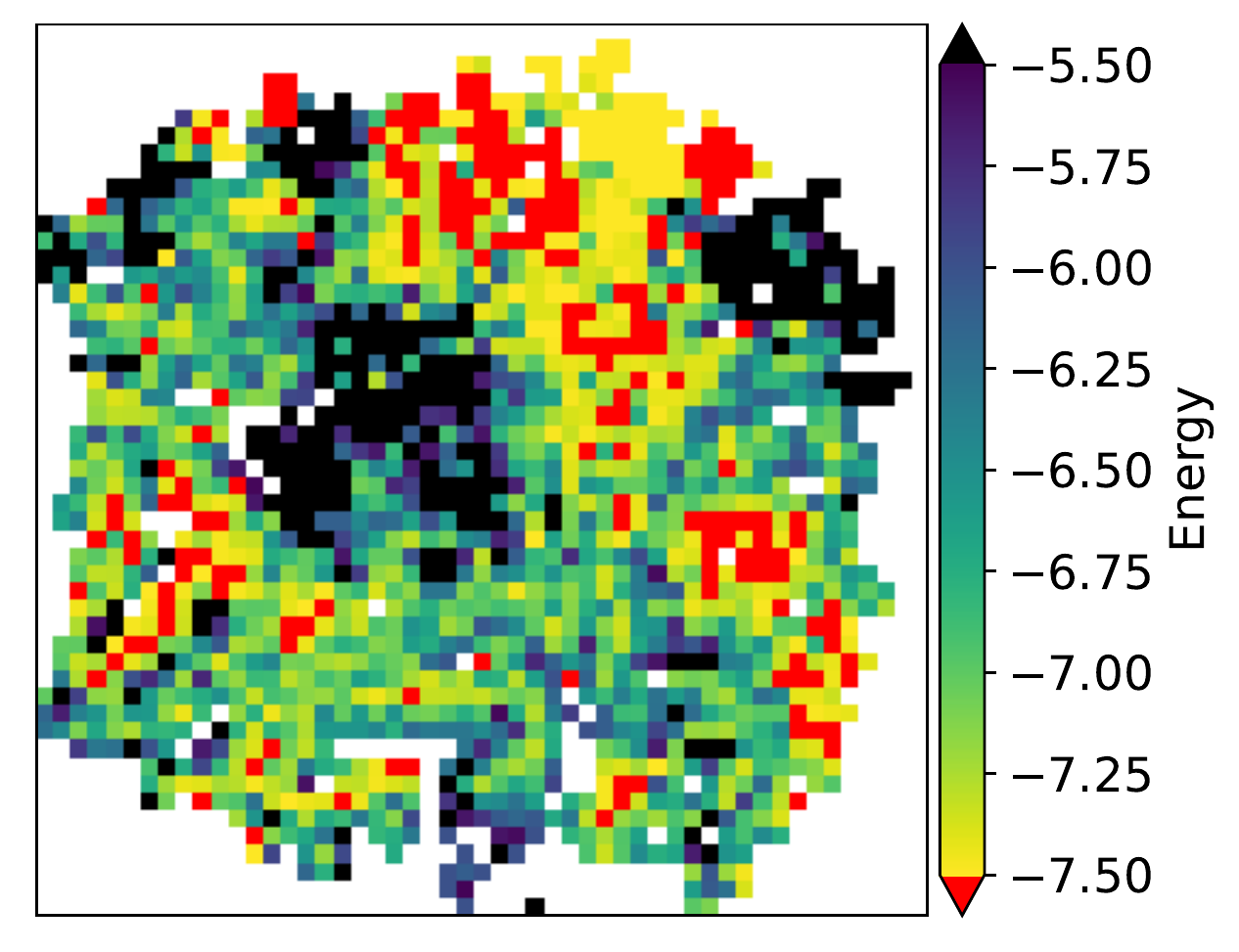}}
\caption{Latent space 2D visualization of the proposed encoding (left) compared to the image encoding (right) \cite{zhang2021neural}  using t-SNE \cite{Laurens2008Visualizing}. Color denotes the VQE energy using the quantum circuits. We randomly sample 50,000 circuits and average the energy in each small area. The areas with average energy lower than -7.5 are marked in red.}
\label{Fig:visualization}
\end{figure}

We visualize the latent spaces of the proposed graph encoding (left) compared to the image encoding (right) \cite{zhang2021neural} in Figure \ref{Fig:visualization}. A well-known visualization tool t-SNE \cite{Laurens2008Visualizing} is used to    convert the high-dimensional encoding of the quantum circuit into two-dimensional data.
The areas with average energy lower than -7.5 are marked in red to highlight the high-performance circuits. For the proposed graph encoding, most of the high-performance circuits (marked in red) concentrate in the lower left region.
The graph encoding maps quantum circuits with similar performances to the same region and has a smooth encoding space with respect to the circuits' performance, which is very conducive to the learning of the predictor.
For the image encoding, the transition of energies is less smooth compared to the graph encoding, and the quantum circuits with low energies (marked in red) scatter over multiple areas.

\subsubsection{Result of the binary classifier}
In this section, we show the results of the binary classifier, which is used to filter out quantum circuits with poor performances. The training set of the binary classifier consists of 400 quantum circuits.
The trained classifier is used to filter out poor circuits from 50,000 quantum circuits drawn from the search space randomly.
Table \ref{tab:res-stage1} shows the number of the remained circuits ($n_{\text{rest}}$), the number of circuits whose ground-truth energies are lower than -7.7 ($n_{e<-7.7}$) in the remained circuits and the density of high-performance circuits after filtering ($n_{e<-7.7}/n_{\text{rest}}$).
The classifier based on graph encoding (GQAS) retains 27 high-performance circuits, nearly twice as many as the one based on image encoding (PQAS), which indicates that it can better filter out poor circuits and improve the efficiency of QAS.
It should be noted that the number of trainable parameters in GQAS is 571, which is only 15\% of that in PQAS's classifier implemented by CNN.
The classifier based on unsupervised representation learning GSQAS(URL) achieves similar performances as GQAS.
By using the pre-training and fine-tuning scheme, GSQAS(PF) filters out more circuits than the other methods and its density of high-performance circuits is similar to GQAS.
\begin{table}
\centering
\caption{Result of the binary classifier. The simulation runs 50 times independently and the average result is reported. PQAS denotes the binary classifier based on image encoding \cite{zhang2021neural} while GQAS is the one based on graph encoding.
GSQAS(URL) and GSQAS(PF) denote the binary classifiers with different self-supervised learning schemes, \ie unsupervised representation learning (URL) and pre-training and fine-tuning (PF).}
\label{tab:res-stage1}
\setlength{\tabcolsep}{3pt}
\centering
\begin{tabular}{cccc}
\hline
Classifier& $n_{e<-7.7}$& $n_{\text{rest}}$ & $n_{e<-7.7}/n_{\text{rest}}   $\\
\hline
PQAS \cite{zhang2021neural}& 15&11259&0.0013\\
GQAS & 27&13706&0.0020\\
GSQAS(URL) & 27&13750&0.0020\\
GSQAS(PF) & 20&10861&0.0018\\
\hline
\end{tabular}
\end{table}

\subsubsection{Result of the predictor}
In the second screening stage, the predictor is a regression model, which estimates the energies of the remained quantum circuits after the first screening stage.
When the energy is lower than -7.7, we assume VQE has found the circuit to achieve the ground-state energy of the TFIM model.
We analyze the performance using three settings: (1) the average probability of achieving the ground-state energy of the TFIM model using different numbers of training circuits to learn a predictor; (2) the average probability of achieving the ground-state energy using different numbers of candidate circuits in the final selection phase; (3) the average number of candidate circuits required to achieve the ground-state energy.

\begin{figure}
\centering
\includegraphics[width=0.6\textwidth]{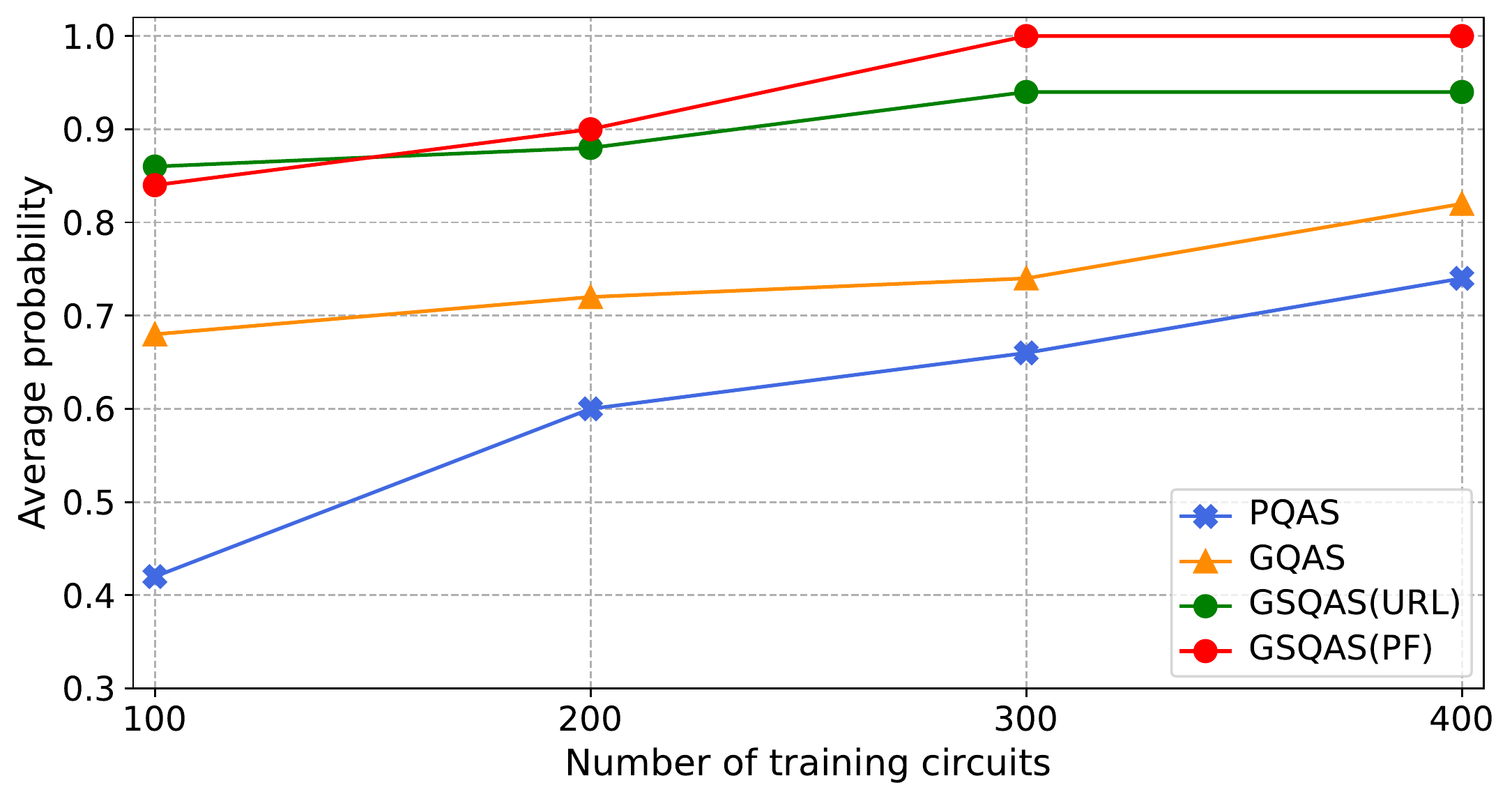}\\
\caption{Average probabilities of achieving the ground-state energy of  the TFIM model over 50 independent runs by using different QAS algorithms with different numbers of training circuits. The number of candidate circuits is fixed at 400. The horizontal axis is the number of training circuits for the predictors.}
\label{Fig:successprobabilityVStrain}
\end{figure}

In the first setting, we show the average probabilities of VQE achieving the ground-state energy over 50 independent runs with respect to the increasing numbers of training circuits in Figure \ref{Fig:successprobabilityVStrain}.
We set the number of candidate circuits in the final selection step as 400.
Using more training samples can increase the probability of achieving the ground-state energy, but it comes with a higher computational cost.
GQAS achieves 8\%-26\% improvements compared to PQAS, which demonstrates the effectiveness of the graph encoding scheme. The number of trainable parameters in GQAS is 571, which is only 33\% of that in PQAS's predictor implemented by LSTM.
The predictors with self-supervised learning, \ie GSQAS(URL) and GSQAS(PF), achieve 12\%-26\% improvements compared to their supervised counterpart GQAS.
GSQAS(URL) and GSQAS(PF) trained with 100 training samples outperform their counterpart GQAS using 400 training samples, which indicates that SSL can largely reduce the number of training samples required to train a high-performance predictor by using the informative knowledge extracted from the unlabeled circuits.
When the number of training samples is 300, GSQAS(PF) has a 100\% probability to achieve the ground-state energy, which is much higher than the other methods.

\begin{figure}
\centering
\includegraphics[width=0.6\textwidth]{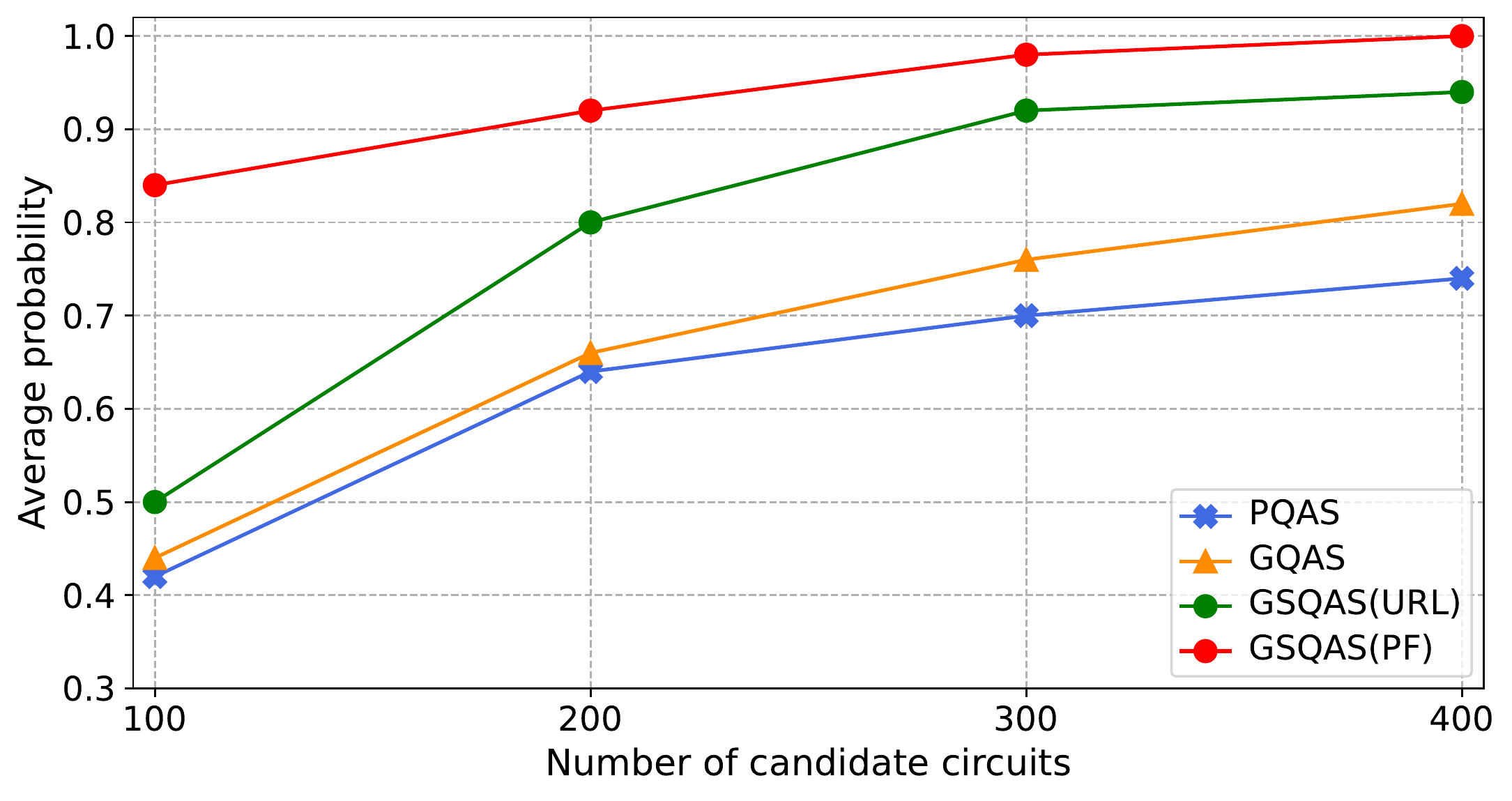}\\
\caption{Average probabilities of achieving the ground-state energy of the TFIM model over 50 independent runs by using different QAS algorithms with different numbers of candidate circuits.
The number of labeled training circuits is fixed at 400.
}
\label{Fig:successprobabilityVSQuery}
\end{figure}

In the second setting, we train the predictor with 400 labeled circuits and use different numbers of candidate circuits in the final selection phase. The performance comparison is shown in Figure \ref{Fig:successprobabilityVSQuery}.
The probability increases with the number of candidate circuits as calculating the ground-truth energies of more circuits will increase the probability to achieve the ground-state energy.
The predictor with graph encoding (GQAS) outperforms the one with image encoding (PQAS).
The predictors trained with self-supervised learning have much higher probabilities to achieve the ground-state energy than its supervised counterpart GQAS.
GSQAS(PF) achieves the best performance across all different numbers of candidate circuits and is still reliable even when the number of candidate circuits is 100. GSQAS(PF) using 100 candidate circuits achieves a higher probability than GQAS using 400 candidate circuits.

\begin{figure}
\centering
\includegraphics[width=0.6\textwidth]{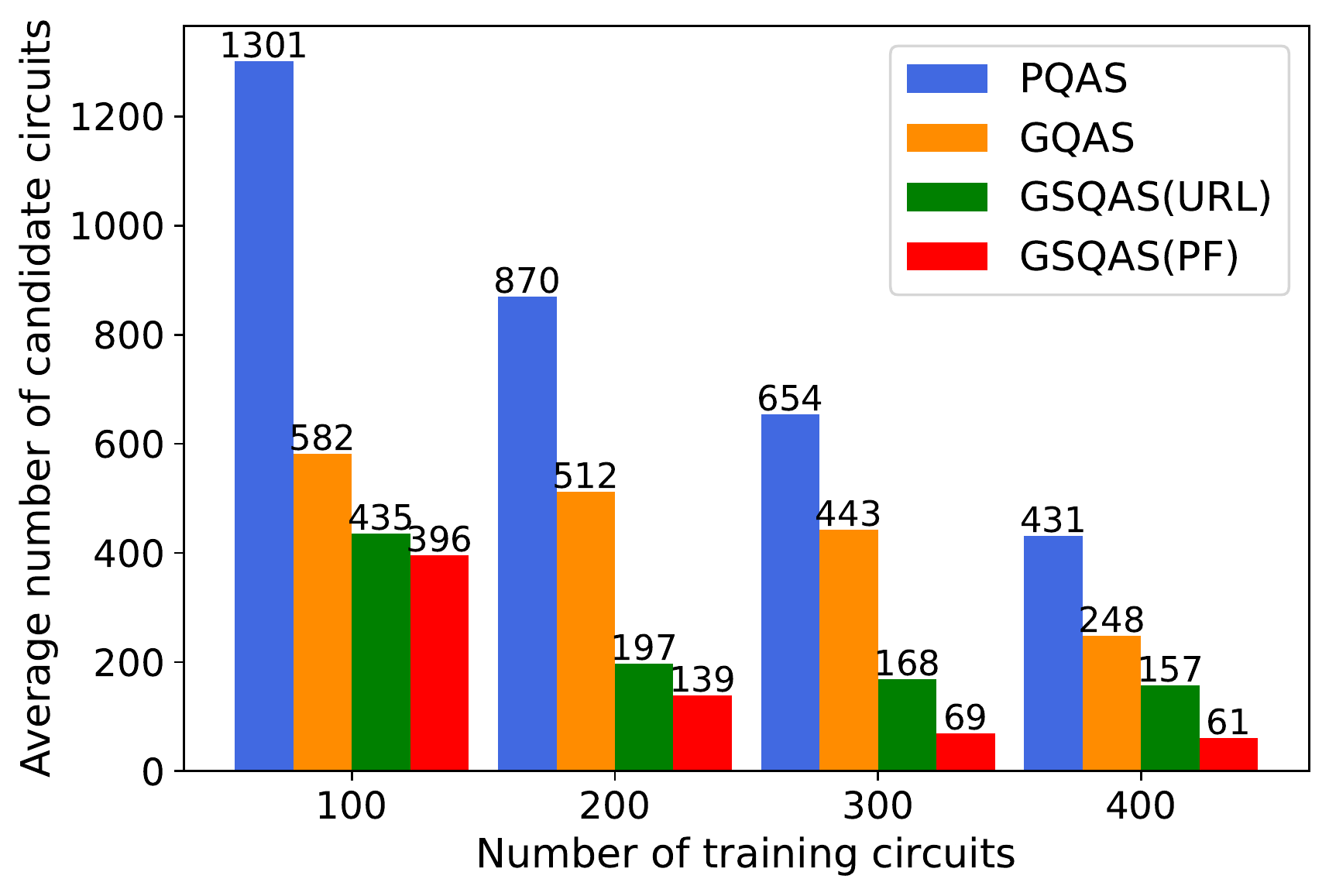}\\
\caption{Average numbers of candidate circuits required to achieve the ground-state energy of the TFIM model over 50 independent runs.}
\label{Fig:queryNum}
\end{figure}

In the last setting, we do not restrict the number of candidate circuits in the final selection step and keep calculating the ground-truth performances of candidate circuits until the ground-state energy of the TFIM model is found. Figure \ref{Fig:queryNum} shows the number of candidate circuits required to achieve the ground-state energy.
Fewer candidate circuits indicate that the predictor is more accurate at estimating the performance of quantum circuits.
With the proposed encoding algorithm, GQAS learns a more accurate predictor and thus largely decreases the number of required candidate circuits compared to PQAS.
With unsupervised representation learning, GSQAS(URL) reduces 25\%-62\% of candidate circuits compared to GQAS. GSQAS(PF) achieves the best performance, requiring only 16\%-68\% of candidate circuits required by GQAS.
On average, GSQAS(URL) and GSQAS(PF) only require 1/2 and 1/3 of candidate circuits required by GQAS to achieve the ground-state energy, which indicates the predictors trained with self-supervised learning are much more accurate than their supervised counterpart.
The number of labeled circuits required by the predictor-based QAS is the summation of the number of training and candidate circuits.
The minimum number of required labeled circuits is 339, which is achieved by GSQAS(PF) using 200 training circuits.
200 seems to be the best size for training circuits as increasing the number of training circuits will increase the total number of required labeled circuits and thus increase the computational complexity.

\subsection{Variational quantum classifier for quantum state classification}
In this section, we evaluate the proposed method on a quantum state classification task.
Ref. \cite{schatzki2021entangled} provided an entangled dataset, containing quantum states with different amounts of concentratable entanglement (CE).
We use the 8-qubit quantum states with CE=0.15 and CE=0.45 to construct a binary classification problem.
A variational quantum classifier implemented by a parameterized quantum circuit(PQC) is trained for this task on the quantum dataset, and QAS searches for a suitable PQC to achieve high accuracy on the test set.
Similar to the variational quantum classifier in Ref. \cite{zhang2021neural}, the primitive set of quantum gates contains $R_x$, $R_y$, $R_z$, $XX$, $YY$, $ZZ$ and pSWAP. pSWAP denotes the parameterized SWAP gate $\text{pSWAP} = e^{-i\theta\text{SWAP}/2}$, where $\text{SWAP}_{12} = \sigma_0^1\sigma_0^2+\sigma_x^1\sigma_x^2+\sigma_y^1\sigma_y^2+\sigma_z^1\sigma_z^2$.
The number of quantum gates and the maximum depth of the circuit are set to $32$ and $7$, respectively.
According to this setting, we have $\newMat{X} \in \mathbb{R}^{34\times 17}$, $\newMat{A} \in \mathbb{R}^{34\times 34}$. We set $l=F$ and thus have $\newMat{Z} \in \mathbb{R}^{34\times 17}$.
Similar to Ref. \cite{zhang2021neural}, we use a layerwise pipeline to generate quantum circuits.

The variational quantum classifier is constructed by a PQC and a simple feedforward neural network with a hidden layer of 16 neurons.
The activation functions of the hidden layer and the output layer are ReLU and Softmax, respectively.
The PQC first takes a quantum state $|\psi \rangle$ as input and then we collect the measurement expectation values of all qubits on the output state $|\psi'\rangle$ of PQC to get a feature vector $\vct m = (m_1,...,m_n)$ by
\begin{equation}
	m_i = \langle \psi'|\sigma_z^i|\psi'\rangle,
	\label{eq:ClassificationMeasure}
\end{equation}
where $\sigma_z^i$ is the observable being measured on the $i$-th qubit. The feedforward neural network takes the feature vector $\vct m$ as input and outputs the label of the quantum state. We use the cross-entropy loss as the loss function.
We collect 400 quantum states with CE=0.15 and 400 quantum states with CE=0.45. Half of the quantum states are randomly selected as the training data of the variational quantum classifier and the rest are test data to evaluate the performance of the classifier.

\begin{figure}
\centering
\includegraphics[width=0.6\textwidth]{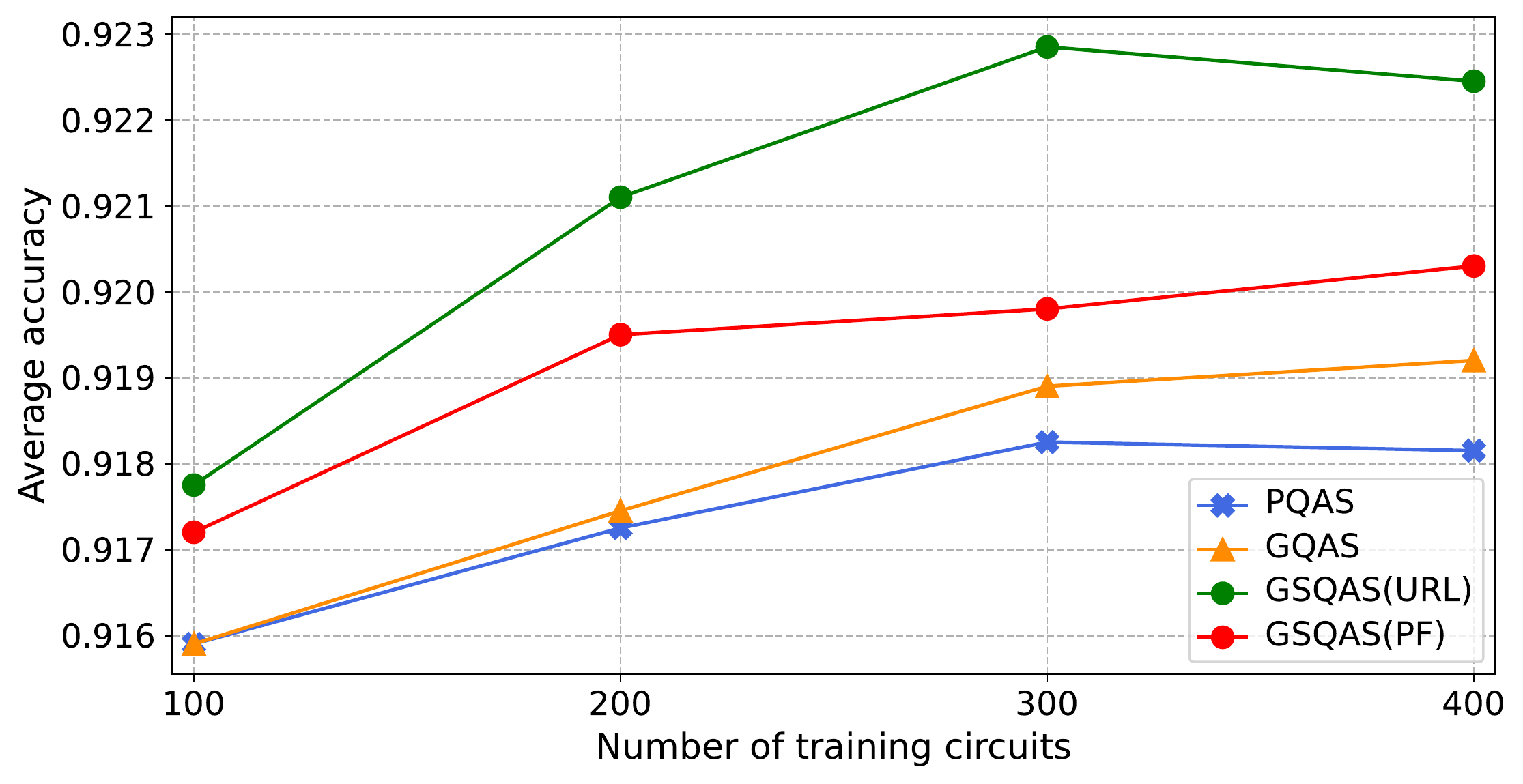}\\
\caption{Average classification accuracies of the variational quantum classifier designed by different QAS algorithms over 50 independent runs by using different numbers of training circuits.}
\label{Fig:accuracyVStrain}
\end{figure}

By setting the number of candidate circuits in the final selection step of predictor-based QAS as 1000, Figure \ref{Fig:accuracyVStrain} shows the average classification accuracy of the variational quantum classifier designed by the predictor-based QAS algorithms using different sizes of training circuits.
Similar to the results of VQE, GQAS achieves higher accuracy than PQAS, which demonstrates the effectiveness of the graph encoding scheme.
The number of trainable parameters in GQAS is 631, which is only 29\% of that in PQAS's predictor implemented by LSTM.
The predictors have a significant improvement in the classification accuracies by using the self-supervised learning methods, especially for GSQAS(URL).
By using 200 training circuits, GSQAS(URL) and GSQAS(PF) achieve higher accuracies than their supervised counterpart GQAS using 400 training circuits.

\begin{figure}
\centering
\includegraphics[width=0.6\textwidth]{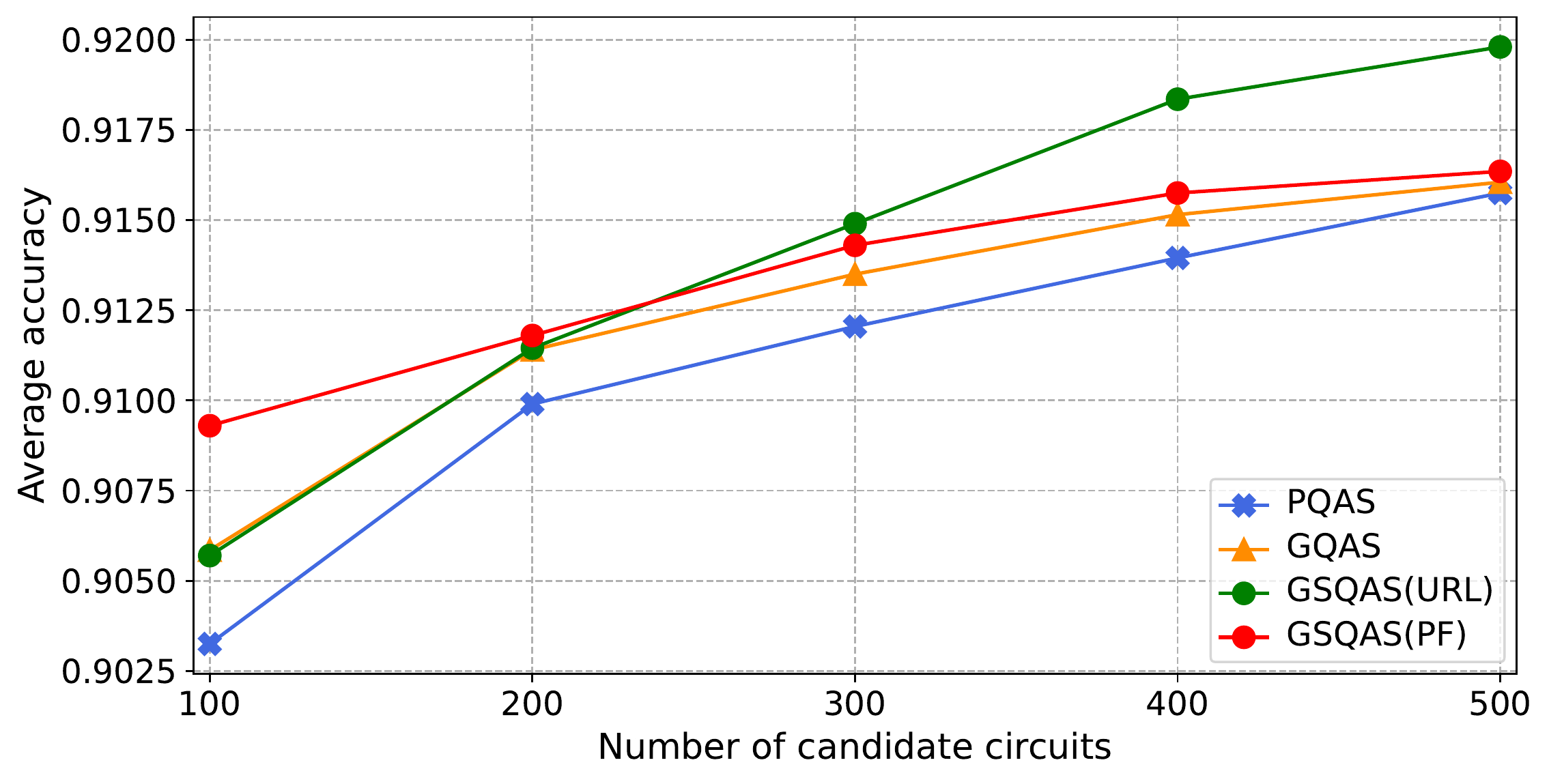}\\
\caption{Average classification accuracies of the variational quantum classifier designed by different QAS algorithms over 50 independent runs by using different numbers of candidate circuits.}
\label{Fig:accuracyVSquery}
\end{figure}

We also train the predictor with 400 training samples and show the average classification accuracies of the variational quantum classifier designed by predictor-based QAS with different numbers of candidate circuits in Figure \ref{Fig:accuracyVSquery}.
GQAS achieves better performances than PQAS in all cases.
GSQAS(PF) achieves the highest accuracy when the numbers of candidate circuits are small, \ie 100 and 200. As the number of candidate circuits increases, GSQAS(URL) has a significant improvement and achieves the highest accuracy.

\section{Conclusion}
In this paper, we propose GSQAS to improve the efficiency of predictor-based QAS by using graph self-supervised learning. More specifically, we first propose a graph encoding scheme for quantum circuits, which is suitable for self-supervised learning and can better encode the spatial topology information. In order to generate an informative representation of quantum circuits, a pretext task is designed to pre-train an encoder on a large number of unlabeled quantum circuits using a reconstruction loss.
Two self-supervised schemes, \ie unsupervised representation learning (URL) and pre-training and fine-tuning (PF) are adopted to train the downstream predictor.
The simulations on variational quantum eigensolver and variational quantum classifier illustrate the superiority of the proposed graph encoding scheme and the self-supervised pre-training.
Compared with the purely supervised predictor, the predictors based on URL and PF achieve considerable or superior performance with several times less labeled training samples.
GSQAS(URL) achieves the best performance in the VQE task while GSQAS(PF) has the highest classification accuracy in the VQC task.
In future work, we will design other pretext tasks to pre-train the encoder and apply GSQAS to other variational quantum algorithms, \eg quantum compiling and quantum approximate optimization algorithms.

\section*{Acknowledgements}
This work is supported by Guangdong Basic and Applied Basic Research Foundation (No.\,2022A1515140116, 2021A1515012138, 2021A1515012639, 2021A1515011985), Key Platform, Research Project of Education Department of Guangdong Province (No.\,2020KTSCX132), Key Research Project of Universities in Guangdong Province (No.\,2019KZDXM007), National Natural Science Foundation of China (No.\, 61972091).  SZ was supported by  the Major Key Project of PCL.

\section*{References}
\bibliographystyle{unsrt}
\bibliography{bibDB}

\begin{thebibliography}{10}

\bibitem{peruzzo2014variational}
Alberto Peruzzo, Jarrod McClean, Peter Shadbolt, Man-Hong Yung, Xiao-Qi Zhou,
  Peter~J Love, Al{\'a}n Aspuru-Guzik, and Jeremy~L O’brien.
\newblock A variational eigenvalue solver on a photonic quantum processor.
\newblock {\em Nature Communications}, 5(1):1--7, 2014.

\bibitem{cerezo2021variational}
Marco Cerezo, Andrew Arrasmith, Ryan Babbush, Simon~C Benjamin, Suguru Endo,
  Keisuke Fujii, Jarrod~R McClean, Kosuke Mitarai, Xiao Yuan, Lukasz Cincio,
  et~al.
\newblock Variational quantum algorithms.
\newblock {\em Nature Reviews Physics}, pages 1--20, 2021.

\bibitem{higgott2019variational}
Oscar Higgott, Daochen Wang, and Stephen Brierley.
\newblock Variational quantum computation of excited states.
\newblock {\em Quantum}, 3:156, 2019.

\bibitem{kandala2017hardware}
Abhinav Kandala, Antonio Mezzacapo, Kristan Temme, Maika Takita, Markus Brink,
  Jerry~M Chow, and Jay~M Gambetta.
\newblock Hardware-efficient variational quantum eigensolver for small
  molecules and quantum magnets.
\newblock {\em Nature}, 549(7671):242--246, 2017.

\bibitem{jones2019variational}
Tyson Jones, Suguru Endo, Sam McArdle, Xiao Yuan, and Simon~C Benjamin.
\newblock Variational quantum algorithms for discovering {H}amiltonian spectra.
\newblock {\em Physical Review A}, 99(6):062304, 2019.

\bibitem{moussa2020quantum}
Charles Moussa, Henri Calandra, and Vedran Dunjko.
\newblock To quantum or not to quantum: towards algorithm selection in
  near-term quantum optimization.
\newblock {\em Quantum Science and Technology}, 5(4):044009, 2020.

\bibitem{mcardle2019variational}
Sam McArdle, Tyson Jones, Suguru Endo, Ying Li, Simon~C Benjamin, and Xiao
  Yuan.
\newblock Variational ansatz-based quantum simulation of imaginary time
  evolution.
\newblock {\em npj Quantum Information}, 5(1):1--6, 2019.

\bibitem{yao2021adaptive}
Yong-Xin Yao, Niladri Gomes, Feng Zhang, Cai-Zhuang Wang, Kai-Ming Ho, Thomas
  Iadecola, and Peter~P Orth.
\newblock Adaptive variational quantum dynamics simulations.
\newblock {\em PRX Quantum}, 2(3):030307, 2021.

\bibitem{he2021variational}
Zhimin He, Lvzhou Li, Shenggen Zheng, Yongyao Li, and Haozhen Situ.
\newblock Variational quantum compiling with double q-learning.
\newblock {\em New Journal of Physics}, 23(3):033002, 2021.

\bibitem{khatri2019quantum}
Sumeet Khatri, Ryan LaRose, Alexander Poremba, Lukasz Cincio, Andrew~T
  Sornborger, and Patrick~J Coles.
\newblock Quantum-assisted quantum compiling.
\newblock {\em Quantum}, 3:140, 2019.

\bibitem{farhi2014quantum}
Edward Farhi, Jeffrey Goldstone, and Sam Gutmann.
\newblock A quantum approximate optimization algorithm.
\newblock {\em arXiv:1411.4028}, 2014.

\bibitem{shi2022parameterized}
Jinjing Shi, Wenxuan Wang, Xiaoping Lou, Shichao Zhang, and Xuelong Li.
\newblock Parameterized {H}amiltonian learning with quantum circuit.
\newblock {\em IEEE Transactions on Pattern Analysis and Machine Intelligence},
  2022.

\bibitem{beer2020training}
Kerstin Beer, Dmytro Bondarenko, Terry Farrelly, Tobias~J Osborne, Robert
  Salzmann, Daniel Scheiermann, and Ramona Wolf.
\newblock Training deep quantum neural networks.
\newblock {\em Nature Communications}, 11(1):1--6, 2020.

\bibitem{li2020quantumCNN}
Yaochong Li, Rigui Zhou, Ruqing Xu, Jia Luo, and Wenwen Hu.
\newblock A quantum deep convolutional neural network for image recognition.
\newblock {\em Quantum Science and Technology}, 5(4):044003, 2020.

\bibitem{niu2022entangling}
Murphy~Yuezhen Niu, Alexander Zlokapa, Michael Broughton, Sergio Boixo, Masoud
  Mohseni, Vadim Smelyanskyi, and Hartmut Neven.
\newblock Entangling quantum generative adversarial networks.
\newblock {\em Physical Review Letters}, 128(22):220505, 2022.

\bibitem{situ2020quantum}
Haozhen Situ, Zhimin He, Yuyi Wang, Lvzhou Li, and Shenggen Zheng.
\newblock Quantum generative adversarial network for generating discrete
  distribution.
\newblock {\em Information Sciences}, 538:193--208, 2020.

\bibitem{shi2023quantum}
Jinjing Shi, Yongze Tang, Yuhu Lu, Yanyan Feng, Ronghua Shi, and Shichao Zhang.
\newblock Quantum circuit learning with parameterized boson sampling.
\newblock {\em IEEE Transactions on Knowledge \& Data Engineering},
  35(02):1965--1976, 2023.

\bibitem{mcclean2018barren}
Jarrod~R McClean, Sergio Boixo, Vadim~N Smelyanskiy, Ryan Babbush, and Hartmut
  Neven.
\newblock Barren plateaus in quantum neural network training landscapes.
\newblock {\em Nature Communications}, 9(1):1--6, 2018.

\bibitem{marrero2021entanglement}
Carlos~Ortiz Marrero, M{\'a}ria Kieferov{\'a}, and Nathan Wiebe.
\newblock Entanglement-induced barren plateaus.
\newblock {\em PRX Quantum}, 2(4):040316, 2021.

\bibitem{cerezo2021cost}
Marco Cerezo, Akira Sone, Tyler Volkoff, Lukasz Cincio, and Patrick~J Coles.
\newblock Cost function dependent barren plateaus in shallow parametrized
  quantum circuits.
\newblock {\em Nature Communications}, 12(1):1--12, 2021.

\bibitem{pesah2021absence}
Arthur Pesah, Marco Cerezo, Samson Wang, Tyler Volkoff, Andrew~T Sornborger,
  and Patrick~J Coles.
\newblock Absence of barren plateaus in quantum convolutional neural networks.
\newblock {\em Physical Review X}, 11(4):041011, 2021.

\bibitem{sharma2022trainability}
Kunal Sharma, Marco Cerezo, Lukasz Cincio, and Patrick~J Coles.
\newblock Trainability of dissipative perceptron-based quantum neural networks.
\newblock {\em Physical Review Letters}, 128(18):180505, 2022.

\bibitem{meng2021quantum}
Fan-Xu Meng, Ze-Tong Li, Xu-Tao Yu, and Zai-Chen Zhang.
\newblock Quantum circuit architecture optimization for variational quantum
  eigensolver via monto carlo tree search.
\newblock {\em IEEE Transactions on Quantum Engineering}, 2:1--10, 2021.

\bibitem{chivilikhin2020mog}
D~Chivilikhin, A~Samarin, V~Ulyantsev, I~Iorsh, AR~Oganov, and O~Kyriienko.
\newblock Mog-vqe: Multiobjective genetic variational quantum eigensolver.
\newblock {\em arXiv:2007.04424}, 2020.

\bibitem{cincio2021machine}
Lukasz Cincio, Kenneth Rudinger, Mohan Sarovar, and Patrick~J Coles.
\newblock Machine learning of noise-resilient quantum circuits.
\newblock {\em PRX Quantum}, 2(1):010324, 2021.

\bibitem{grimsley2019adaptive}
Harper~R Grimsley, Sophia~E Economou, Edwin Barnes, and Nicholas~J Mayhall.
\newblock An adaptive variational algorithm for exact molecular simulations on
  a quantum computer.
\newblock {\em Nature Communications}, 10(1):1--9, 2019.

\bibitem{li2020quantum}
Li~Li, Minjie Fan, Marc Coram, Patrick Riley, Stefan Leichenauer, et~al.
\newblock Quantum optimization with a novel {Gibbs} objective function and
  ansatz architecture search.
\newblock {\em Physical Review Research}, 2(2):023074, 2020.

\bibitem{zhang2022differentiable}
Shi-Xin Zhang, Chang-Yu Hsieh, Shengyu Zhang, and Hong Yao.
\newblock Differentiable quantum architecture search.
\newblock {\em Quantum Science and Technology}, 7(4):045023, 2022.

\bibitem{zhang2021neural}
Shi-Xin Zhang, Chang-Yu Hsieh, Shengyu Zhang, and Hong Yao.
\newblock Neural predictor based quantum architecture search.
\newblock {\em Machine Learning: Science and Technology}, 2(4):045027, 2021.

\bibitem{du2022quantum}
Yuxuan Du, Tao Huang, Shan You, Min-Hsiu Hsieh, and Dacheng Tao.
\newblock Quantum circuit architecture search for variational quantum
  algorithms.
\newblock {\em npj Quantum Information}, 8(1):1--8, 2022.

\bibitem{he2022AQT}
Zhimin He, Chuangtao Chen, Lvzhou Li, Shenggen Zheng, and Haozhen Situ.
\newblock Quantum architecture search with meta-learning.
\newblock {\em Advanced Quantum Technologies}, 5(8):2100134, 2022.

\bibitem{he2022search}
Zhimin He, Junjian Su, Chuangtao Chen, Minghua Pan, and Haozhen Situ.
\newblock Search space pruning for quantum architecture search.
\newblock {\em The European Physical Journal Plus}, 137(4):491, 2022.

\bibitem{linghu2022quantum}
Kehuan Linghu, Yang Qian, Ruixia Wang, Meng-Jun Hu, Zhiyuan Li, Xuegang Li,
  Huikai Xu, Jingning Zhang, Teng Ma, Peng Zhao, et~al.
\newblock Quantum circuit architecture search on a superconducting processor.
\newblock {\em arXiv:2201.00934}, 2022.

\bibitem{lu2021markovian}
Zhide Lu, Pei-Xin Shen, and Dong-Ling Deng.
\newblock Markovian quantum neuroevolution for machine learning.
\newblock {\em Physical Review Applied}, 16(4):044039, 2021.

\bibitem{ostaszewski2021structure}
Mateusz Ostaszewski, Edward Grant, and Marcello Benedetti.
\newblock Structure optimization for parameterized quantum circuits.
\newblock {\em Quantum}, 5:391, 2021.

\bibitem{huang2022robust}
Yuhan Huang, Qingyu Li, Xiaokai Hou, Rebing Wu, Man-Hong Yung, Abolfazl Bayat,
  and Xiaoting Wang.
\newblock Robust resource-efficient quantum variational ansatz through an
  evolutionary algorithm.
\newblock {\em Physical Review A}, 105(5):052414, 2022.

\bibitem{rattew2019domain}
Arthur~G Rattew, Shaohan Hu, Marco Pistoia, Richard Chen, and Steve Wood.
\newblock A domain-agnostic, noise-resistant, hardware-efficient evolutionary
  variational quantum eigensolver.
\newblock {\em arXiv:1910.09694}, 2019.

\bibitem{zhang20}
Yuan-Hang Zhang, Pei-Lin Zheng, Yi~Zhang, and Dong-Ling Deng.
\newblock Topological quantum compiling with reinforcement learning.
\newblock {\em Physical Review Letters}, 125(17):170501, 2020.

\bibitem{ostaszewski2021reinforcement}
Mateusz Ostaszewski, Lea~M Trenkwalder, Wojciech Masarczyk, Eleanor Scerri, and
  Vedran Dunjko.
\newblock Reinforcement learning for optimization of variational quantum
  circuit architectures.
\newblock {\em Advances in Neural Information Processing Systems}, 2021.

\bibitem{kuo2021quantum}
En-Jui Kuo, Yao-Lung~L Fang, and Samuel Yen-Chi Chen.
\newblock Quantum architecture search via deep reinforcement learning.
\newblock {\em arXiv:2104.07715}, 2021.

\bibitem{wang2022quantumnas}
Hanrui Wang, Yongshan Ding, Jiaqi Gu, Yujun Lin, David~Z Pan, Frederic~T Chong,
  and Song Han.
\newblock Quantum{NAS}: Noise-adaptive search for robust quantum circuits.
\newblock In {\em International Symposium on High-Performance Computer
  Architecture (HPCA)}, pages 692--708. IEEE, 2022.

\bibitem{he2020momentum}
Kaiming He, Haoqi Fan, Yuxin Wu, Saining Xie, and Ross Girshick.
\newblock Momentum contrast for unsupervised visual representation learning.
\newblock In {\em Proceedings of the IEEE/CVF conference on computer vision and
  pattern recognition}, pages 9729--9738, 2020.

\bibitem{chen2020simple}
Ting Chen, Simon Kornblith, Mohammad Norouzi, and Geoffrey Hinton.
\newblock A simple framework for contrastive learning of visual
  representations.
\newblock In {\em International conference on machine learning}, pages
  1597--1607. PMLR, 2020.

\bibitem{wei2022masked}
Chen Wei, Haoqi Fan, Saining Xie, Chao-Yuan Wu, Alan Yuille, and Christoph
  Feichtenhofer.
\newblock Masked feature prediction for self-supervised visual pre-training.
\newblock In {\em Proceedings of the IEEE/CVF Conference on Computer Vision and
  Pattern Recognition}, pages 14668--14678, 2022.

\bibitem{he2022masked}
Kaiming He, Xinlei Chen, Saining Xie, Yanghao Li, Piotr Doll{\'a}r, and Ross
  Girshick.
\newblock Masked autoencoders are scalable vision learners.
\newblock In {\em Proceedings of the IEEE/CVF Conference on Computer Vision and
  Pattern Recognition}, pages 16000--16009, 2022.

\bibitem{devlin2018bert}
Jacob Devlin, Ming-Wei Chang, Kenton Lee, and Kristina Toutanova.
\newblock Bert: Pre-training of deep bidirectional transformers for language
  understanding.
\newblock {\em arXiv:1810.04805}, 2018.

\bibitem{yang2019xlnet}
Zhilin Yang, Zihang Dai, Yiming Yang, Jaime Carbonell, Russ~R Salakhutdinov,
  and Quoc~V Le.
\newblock Xlnet: Generalized autoregressive pretraining for language
  understanding.
\newblock {\em Advances in neural information processing systems}, 32, 2019.

\bibitem{qian2022contentvec}
Kaizhi Qian, Yang Zhang, Heting Gao, Junrui Ni, Cheng-I Lai, David Cox, Mark
  Hasegawa-Johnson, and Shiyu Chang.
\newblock Contentvec: An improved self-supervised speech representation by
  disentangling speakers.
\newblock In {\em International Conference on Machine Learning}, pages
  18003--18017. PMLR, 2022.

\bibitem{choi2021neural}
Hyeong-Seok Choi, Juheon Lee, Wansoo Kim, Jie Lee, Hoon Heo, and Kyogu Lee.
\newblock Neural analysis and synthesis: Reconstructing speech from
  self-supervised representations.
\newblock {\em Advances in Neural Information Processing Systems},
  34:16251--16265, 2021.

\bibitem{velickovic2019deep}
Petar Velickovic, William Fedus, William~L Hamilton, Pietro Li{\`o}, Yoshua
  Bengio, and R~Devon Hjelm.
\newblock Deep graph infomax.
\newblock {\em International Conference on Learning Representations}, 2(3):4,
  2019.

\bibitem{hassani2020contrastive}
Kaveh Hassani and Amir~Hosein Khasahmadi.
\newblock Contrastive multi-view representation learning on graphs.
\newblock In {\em International Conference on Machine Learning}, pages
  4116--4126. PMLR, 2020.

\bibitem{hu2020gpt}
Ziniu Hu, Yuxiao Dong, Kuansan Wang, Kai-Wei Chang, and Yizhou Sun.
\newblock Gpt-gnn: Generative pre-training of graph neural networks.
\newblock In {\em Proceedings of ACM SIGKDD International Conference on
  Knowledge Discovery \& Data Mining}, pages 1857--1867, 2020.

\bibitem{qiu2020gcc}
Jiezhong Qiu, Qibin Chen, Yuxiao Dong, Jing Zhang, Hongxia Yang, Ming Ding,
  Kuansan Wang, and Jie Tang.
\newblock Gcc: Graph contrastive coding for graph neural network pre-training.
\newblock In {\em Proceedings of SIGKDD International Conference on Knowledge
  Discovery \& Data Mining}, pages 1150--1160, 2020.

\bibitem{hu2019strategies}
Weihua Hu, Bowen Liu, Joseph Gomes, Marinka Zitnik, Percy Liang, Vijay Pande,
  and Jure Leskovec.
\newblock Strategies for pre-training graph neural networks.
\newblock {\em arXiv:1905.12265}, 2019.

\bibitem{wang2021self}
Xiao Wang, Nian Liu, Hui Han, and Chuan Shi.
\newblock Self-supervised heterogeneous graph neural network with
  co-contrastive learning.
\newblock In {\em Proceedings of the 27th ACM SIGKDD Conference on Knowledge
  Discovery \& Data Mining}, pages 1726--1736, 2021.

\bibitem{jiang2021pre}
Xunqiang Jiang, Tianrui Jia, Yuan Fang, Chuan Shi, Zhe Lin, and Hui Wang.
\newblock Pre-training on large-scale heterogeneous graph.
\newblock In {\em Proceedings of the 27th ACM SIGKDD Conference on Knowledge
  Discovery \& Data Mining}, pages 756--766, 2021.

\bibitem{lee2022augmentation}
Namkyeong Lee, Junseok Lee, and Chanyoung Park.
\newblock Augmentation-free self-supervised learning on graphs.
\newblock In {\em Proceedings of the AAAI Conference on Artificial
  Intelligence}, volume~36, pages 7372--7380, 2022.

\bibitem{luo2022clear}
Xiao Luo, Wei Ju, Meng Qu, Yiyang Gu, Chong Chen, Minghua Deng, Xian-Sheng Hua,
  and Ming Zhang.
\newblock Clear: Cluster-enhanced contrast for self-supervised graph
  representation learning.
\newblock {\em IEEE Transactions on Neural Networks and Learning Systems},
  2022.

\bibitem{suresh2021adversarial}
Susheel Suresh, Pan Li, Cong Hao, and Jennifer Neville.
\newblock Adversarial graph augmentation to improve graph contrastive learning.
\newblock {\em Advances in Neural Information Processing Systems},
  34:15920--15933, 2021.

\bibitem{lin2022prototypical}
Shuai Lin, Chen Liu, Pan Zhou, Zi-Yuan Hu, Shuojia Wang, Ruihui Zhao, Yefeng
  Zheng, Liang Lin, Eric Xing, and Xiaodan Liang.
\newblock Prototypical graph contrastive learning.
\newblock {\em IEEE Transactions on Neural Networks and Learning Systems},
  2022.

\bibitem{kipf2016variational}
Thomas~N Kipf and Max Welling.
\newblock Variational graph auto-encoders.
\newblock {\em arXiv:1611.07308}, 2016.

\bibitem{KeyuluXu2018HowPA}
Keyulu Xu, Weihua Hu, Jure Leskovec, and Stefanie Jegelka.
\newblock How powerful are graph neural networks.
\newblock In {\em International Conference on Learning Representations}, 2021.

\bibitem{DiederikPKingma14}
Diederik~P. Kingma and Max Welling.
\newblock Auto-encoding variational bayes.
\newblock In {\em International Conference on Learning Representations}, 2014.

\bibitem{zhang2022tensorcircuit}
Shi-Xin Zhang, Jonathan Allcock, Zhou-Quan Wan, Shuo Liu, Jiace Sun, Hao Yu,
  Xing-Han Yang, Jiezhong Qiu, Zhaofeng Ye, Yu-Qin Chen, et~al.
\newblock Tensorcircuit: a quantum software framework for the nisq era.
\newblock {\em arXiv:2205.10091}, 2022.

\bibitem{Laurens2008Visualizing}
Laurens van~der Maaten and Geoffrey~E. Hinton.
\newblock Visualizing data using t-{SNE}.
\newblock {\em Journal of Machine Learning Research}, 2008.

\bibitem{schatzki2021entangled}
Louis Schatzki, Andrew Arrasmith, Patrick~J Coles, and Marco Cerezo.
\newblock Entangled datasets for quantum machine learning.
\newblock {\em arXiv preprint arXiv:2109.03400}, 2021.

\end{thebibliography}

\end{document}